\documentclass[journal]{IEEEtran}
\usepackage{caption}
\usepackage{float}
\usepackage{booktabs}
\usepackage{graphicx}
\usepackage{subfigure}
\usepackage{amssymb}
\usepackage{bm}
\usepackage{bbding}
\usepackage{indentfirst}
\usepackage{enumerate}
\usepackage{hyperref}
\usepackage{multirow}
\usepackage{makecell}  
\usepackage{nicematrix}
\usepackage[ruled]{algorithm2e}
\usepackage{colortbl}
\definecolor{mygray}{gray}{.95}
\definecolor{mygreen}{RGB}{0, 128, 85}

\makeatletter
\newcommand{\removelatexerror}{\let\@latex@error\@gobble}
\makeatother
\makeatletter
\renewcommand{\maketag@@@}[1]{\hbox{\m@th\normalsize\normalfont#1}}%
\makeatother

\hypersetup{
  colorlinks=true,
  citecolor=blue,
  linkcolor=red,
  urlcolor=red,
  anchorcolor=black
  }

\usepackage{tikz,xcolor}
\definecolor{lime}{HTML}{A6CE39}
\DeclareRobustCommand{\orcidicon}{%
	\begin{tikzpicture}
	\draw[lime, fill=lime] (0,0) 
	circle [radius=0.134] 
	node[white] {{\fontfamily{qag}\selectfont \tiny ID}};    \draw[white, fill=white] (-0.0625,0.095) 
	circle [radius=0.007];    \end{tikzpicture}
	\hspace{-2mm}}
\foreach \x in {A, ..., Z}{%
	\expandafter\xdef\csname orcid\x\endcsname{\noexpand\href{https://orcid.org/\csname orcidauthor\x\endcsname}{\noexpand\orcidicon}}
}

\begin{document}

\captionsetup[figure]{name={Fig.},labelsep=period}
\captionsetup[table]{name={TABLE},labelsep=period}

\title{GDSR: Global-Detail Integration through Dual-Branch Network with Wavelet Losses for Remote Sensing Image Super-Resolution}

\author{Qiwei~Zhu\orcidA{},
        Kai~Li\orcidB{},
       Guojing~Zhang\orcidC{}, 
       Xiaoying~Wang\orcidD{},
       Jianqiang~Huang\orcidE{}
       and Xilai~Li\orcidF{} 
       
\thanks{This work was supported in part by the Natural Science Foundation of Qinghai Province (No.2023-ZJ-906M) and the National Natural Science Foundation of China (No.U23A20159, No.62162053, No.42265010). (\emph{Corresponding author: Guojing Zhang; Xiaoying Wang}.)
 }
\thanks{Qiwei Zhu, Guojing Zhang, and Jianqiang Huang are with the School of Computer Technology and Application, Qinghai University, Xining 810016, China (e-mail:
qiweizhu@qhu.edu.cn; zhanggj@qhu.edu.cn; 2011990026@qhu.edu.cn), and also with the Qinghai Provincial Laboratory for Intelligent Computing and Application, Qinghai University, Xining 810016, China.}
\thanks{Kai Li is with the Department of Computer Science and Technology, Tsinghua University, Beijing 100084, China (e-mail: li-k24@mails.tsinghua.edu.cn).}
\thanks{Xiaoying Wang is with the School of Computer and Information Science, Qinghai Institute of Technology, Xining 810018, China (e-mail: wangxy@qhit.edu.cn).}
\thanks{Xilai Li is with the College of Agriculture and Animal Husbandry, Qinghai University, Xining 810016, China (e-mail: 1985990024@qhu.edu.cn).}
}

\markboth{Journal of \LaTeX\ Class Files,~Vol.~14, No.~8, August~2021}%
{Shell \MakeLowercase{\textit{et al.}}: A Sample Article Using IEEEtran.cls for IEEE Journals}

\IEEEpubid{0000--0000/00\$00.00~\copyright~2021 IEEE}

\maketitle
\begin{abstract}
In recent years, deep neural networks, including Convolutional Neural Networks, Transformers, and State Space Models, have achieved significant progress in Remote Sensing Image (RSI) Super-Resolution (SR). However, existing SR methods typically overlook the complementary relationship between global and local dependencies. These methods either focus on capturing local information or prioritize global information, which results in models that are unable to effectively capture both global and local features simultaneously. Moreover, their computational cost becomes prohibitive when applied to large-scale RSIs. To address these challenges, we introduce the novel application of Receptance Weighted Key Value (RWKV) to RSI-SR, which captures long-range dependencies with linear complexity. To simultaneously model global and local features, we propose the Global-Detail dual-branch structure, GDSR, which performs SR by paralleling RWKV and convolutional operations to handle large-scale RSIs. Furthermore, we introduce the Global-Detail Reconstruction Module (GDRM) as an intermediary between the two branches to bridge their complementary roles. In addition, we propose the Dual-Group Multi-Scale Wavelet Loss, a wavelet-domain constraint mechanism via dual-group subband strategy and cross-resolution frequency alignment for enhanced reconstruction fidelity in RSI-SR. Extensive experiments under two degradation methods on several benchmarks, including AID, UCMerced, and RSSRD-QH, demonstrate that GSDR outperforms the state-of-the-art Transformer-based method HAT by an average of 0.09 dB in PSNR, while using only 63\% of its parameters and 51\% of its FLOPs, achieving an inference speed 3.2 times faster. 
\end{abstract}

\section{Introduction}
\begin{figure}[!t]
\centering
\includegraphics[width=3.4in]{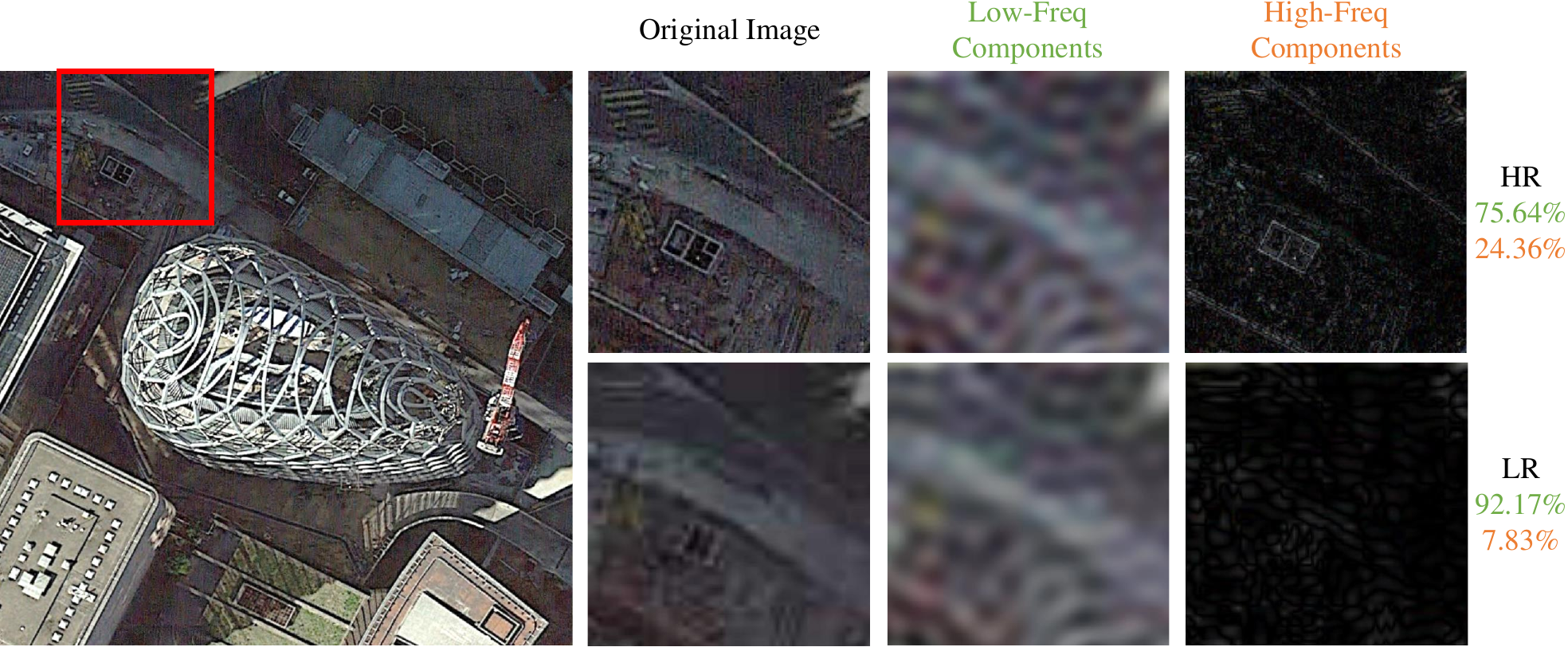}
\captionsetup{font={scriptsize}}
\caption{Example of spatial frequency decomposition between HR and LR RSI.}
\label{fig_1}
\end{figure}

\IEEEPARstart{T}{he} rapid development of remote sensing technologies has resulted in an ever-increasing availability of satellite and aerial images, which play a crucial role in a variety of applications, such as environmental monitoring~\cite{environmentalmonitoring}, urban planning~\cite{li2023machine}, and disaster management~\cite{disastermanagement}. However, the resolution of these remote sensing images is often constrained by the limitations of imaging sensors, atmospheric interference, and satellite bandwidth. Consequently, SR techniques, which seek to reconstruct high-resolution (HR) images from low-resolution (LR) observations, have garnered significant interest from both academic and industrial sectors~\cite{li2020survey,huang2019single}. In particular, remote sensing image super-resolution (RSISR) serves as a pivotal technique for enhancing spatial resolution to extract fine-grained details required for precise geospatial interpretation. However, quantitative frequency analysis highlights a fundamental imbalance in spatial frequency composition between LR and HR observations. As illustrated in Fig.~\ref{fig_1}, LR images predominantly contain low-frequency components, while critical high-frequency features representing geospatial patterns are severely attenuated. This asymmetry is exacerbated by conventional degradation processes that disproportionately suppress high-frequency spatial information, creating reconstruction bottlenecks that standard SR methods struggle to resolve effectively.

\IEEEpubidadjcol

\begin{figure}[!t]
\centering
\includegraphics[width=3.4in]{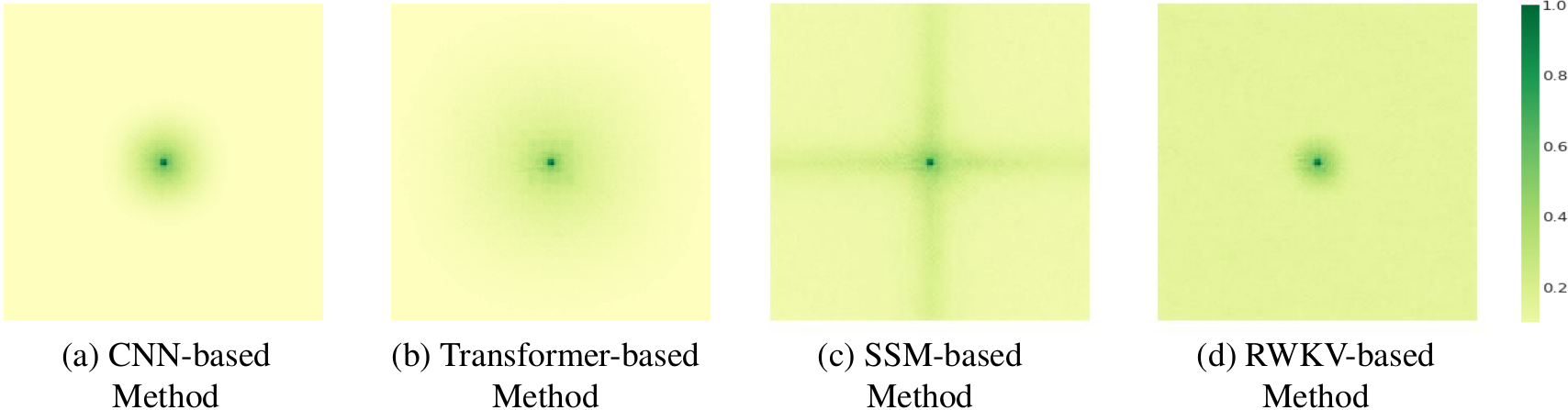}
\captionsetup{font={scriptsize}}
\caption{The ERF~\cite{ERF} visualization for different models with different architectures. A more extensively distributed dark area indicates a larger ERF.}
\label{fig_ERF}
\end{figure}

Traditional SR methods have been dominated by Convolutional Neural Networks (CNNs)~\cite{srcnn, kim2016accurate, srgan, edsr} due to their strong local feature extraction and computational efficiency. However, as depicted in Fig.~\ref{fig_ERF} (a), the limited effective receptive field (ERF)~\cite{ERF} of CNNs inherently restricts their ability to model long-range dependencies. This becomes particularly challenging in large-scale RSIs, where diverse spatial patterns and textures overwhelm the limited modeling capacity of CNN-based methods~\cite{TransENet}. While transformer-based models~\cite{swinir, TransENet, hat} alleviate the ERF limitation through global self-attention mechanisms, their quadratic computational complexity necessitates window-based self-attention that paradoxically shrinks effective receptive field coverage (as depicted in Fig.~\ref{fig_ERF} (b)), and their sequence transformation processes risk losing structural information critical for spatial consistency in RSIs~\cite{gao2021stransfuse}. Recent advances in linear architectures address the computational constraints of Transformers through distinct approaches. Mamba-based models~\cite{mambair, FreMamba} leverage state-space sequences (SSMs)~\cite{gu2023mamba} to achieve linear-complexity long-range dependency modeling, demonstrating performance comparable to Transformers. However, their inherent sequence modeling in SSMs still poses challenges in achieving an optimal ERF for 2D images (as depicted in Fig.~\ref{fig_ERF} (c)), limiting spatial pattern capture in RSIs. Conversely, RWKV-based models combine linear attention mechanisms with localized dependency modeling, outperforming both Transformers and Mamba variants in vision tasks~\cite{peng2024eagle, yuan2024mamba, restorerwkv} and offering efficient long-range modeling with expanded ERF (as depicted in Fig.~\ref{fig_ERF} (d)). Despite pioneering efforts to adopt RWKV for vision tasks, the adaptation of RWKV to the field of RSISR remains unexplored. Crucially, existing approaches spanning CNNs, Transformers, Mamba, and RWKV variants universally neglect the complementary synergy between global attention mechanisms (low-pass filtering) and convolutional operations (high-pass filtering)~\cite{park2022how}, failing to synergize structural context preservation with high-frequency detail recovery essential for precise large-scale RSI reconstruction. This gap underscores the need for architectures that unify efficient global dependency modeling and localized texture refinement tailored to RSISR.

In this paper, we propose a novel approach that integrates global representations and detailed extraction with wavelet losses to enhance the performance of RSISR. Our main contributions can be summarized as follows:

\begin{enumerate}[1)]
\item GDSR (Global-Detail Super-Resolution) is proposed to synergistically unify the RWKV and CNN within a dual-branch framework, where global contextual dependencies and localized spatial features are collaboratively refined for RSISR.
\item A Global–Detail Reconstruction Module (GDRM) is proposed to integrate information from the dual branches efficiently. By merging complementary features, this module enhances the synergy between the branches, thereby significantly improving the overall performance of RSISR tasks.
\item Dual-Group Multi-Scale Wavelet Loss is proposed to unify low-frequency structural subbands and aggregated high-frequency detail subbands through a dual-coefficient grouping strategy, enforcing cross-resolution frequency consistency via hierarchical wavelet decomposition across resolutions within a multi-scale framework, leveraging wavelet-domain frequency priors to augment spatial-domain optimization.
\end{enumerate}

We demonstrate the empirical validity of GDSR by comparing it with various competitive CNN-, GAN-, Transformer-, Mamba-, and hybrid-based models on three RSI benchmark datasets using two degradation methods showing SOTA results. In addition, we conducted extensive ablation experiments (as detailed in Section~\ref{sectionAb}) to investigate the approach of global-detail integration and its effect on RSI-SR, and to examine the effect of Dual-Group Multi-Scale Wavelet Loss. We believe our work will advance efficient large-scale RSI-SR, improving both fidelity and processing speed, while ensuring robustness and practical applicability for real-world remote sensing tasks.

\section{Related Work} \label{section2}
\subsection{Remote Sensing Image Super-resolution}
Recent advances in deep learning have greatly improved RSI-SR, with methods typically falling into CNN-based, Transformer-based, and Mamba-based categories. CNN-based methods~\cite{RDBPN, DSSR, HSENet, HAUNet}, inspired by SRCNN~\cite{srcnn}, have used modules like residual connections and attention mechanisms to enhance performance. However, they are limited by their inability to capture long-range dependencies, which hinders their effectiveness in large-scale RSI data. Transformer-based methods~\cite{TransENet, DSSTSR, SPT} leverage self-attention (SA) to overcome CNN's locality constraints. While standard SA achieves global interaction through pairwise token correlation, its quadratic complexity becomes prohibitive for high-resolution RSIs. Window-based SA~\cite{swinir} reduces computation via local window partitioning, yet critically limits cross-window dependencies required for reconstructing continuous geographic features (e.g., highway systems or river networks) spanning multiple windows. Mamba-based methods emerge as an alternative with linear complexity. MambaIR~\cite{mambair} successfully introduced Mamba~\cite{gu2023mamba} into the SR domain, while FreMamba~\cite{FreMamba} pioneers the exploration of Mamba’s potential for RSI-SR, extending it with frequency analysis. However, SSMs' causal modeling framework, constrained by 1D sequence scanning of 2D spatial data, fundamentally struggles to capture omnidirectional contextual relationships required for coherent reconstruction of geometrically complex remote sensing features (e.g., roundabouts with radial roads). Emerging hybrid architectures combining multiple paradigms have demonstrated significant progress. Recent works~\cite{ConvFormerSR,hybrid1,hybrid2,hybrid3} propose innovative solutions integrating CNN-Transformer feature fusion, U-Net structures with adaptive edge recovery, and shift channel attention-enhanced Transformers, effectively addressing limitations of single-paradigm approaches while maintaining computational efficiency. 

\begin{figure*}[!t]
\centering
\includegraphics[width=6.5in]{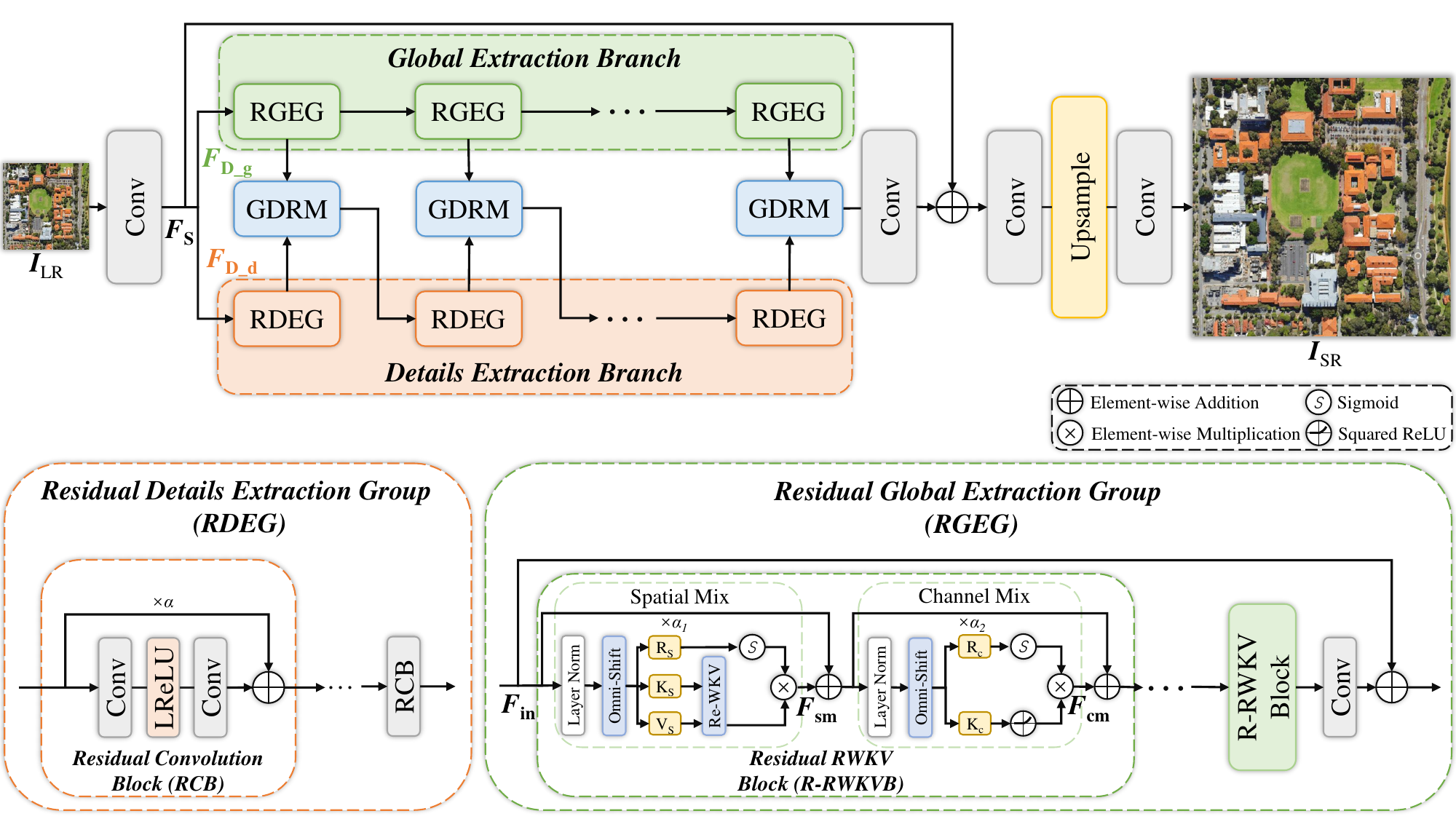}%
\captionsetup{font={scriptsize}}
\caption{Overview of the proposed GDSR.}
\label{fig_Network}
\end{figure*}

\subsection{RWKV in computer vision}
Recent studies~\cite{peng2023rwkv,peng2024eagle, yuan2024mamba} highlight the superiority of RWKV over Transformers and Mamba in NLP tasks due to its linear computational complexity and parallelizable recurrent architecture. In computer vision, Vision-RWKV~\cite{duan2025visionrwkv} adapts RWKV for 2D spatial processing by introducing a bidirectional WKV attention mechanism to capture global dependencies and a quad-directional token shift operation to model local contextual relationships across horizontal and vertical axes. These adaptations preserve the efficiency of RWKV while enabling effective feature learning in images. Subsequent developments further expand RWKV’s applicability. Diffusion-RWKV~\cite{fei2024diffusion} integrates RWKV into diffusion models for image generation, leveraging bidirectional processing to handle patched image inputs through linear-complexity spatial aggregation, thereby eliminating the need for window-based operations. Restore-RWKV~\cite{restorerwkv} enhances 2D dependency modeling through a recurrent WKV (Re-WKV) mechanism that integrates bidirectional and recurrent attention across multiple scan directions, alongside an omnidirectional token shift (Omni-Shift) layer to broaden context aggregation. 

However, existing implementations remain suboptimal for RSI due to distinct domain characteristics. Restore-RWKV’s homogeneous U-Net backbone suits anatomical medical data but falters with RSI’s extreme scale variations and heterogeneous geospatial textures. This paper introduces a dual-branch framework that decouples global feature extraction using RWKV from local feature extraction using CNN, directly addressing RSI-SR-specific challenges unmet by homogeneous architectures.


\subsection{Wavelet Transforms in Deep Learning}
The wavelet transform (WT)~\cite{daubechies1992ten} has been a fundamental signal processing tool since the 1980s and has recently been integrated into neural network architectures. The initial wave of deep learning integration saw wavelet-based architectures like Wavelet-SRNet~\cite{huang2017wavelet}, WDST~\cite{deng2019wavelet} and WDRN~\cite{WDRN} employing wavelet coefficient prediction for SR enhancement. Subsequent innovations deepened this synergy, with WTConv~\cite{finder2025wavelet} introducing trainable wavelet-based layers that dynamically expand convolutional receptive fields, demonstrating cross-task adaptability in computer vision. The integration further evolved through adversarial frameworks, particularly WGSR~\cite{WGSR} that embedded wavelet subband losses within GAN architectures. Moreover,~\cite{Waveletloss} extended this strategy by integrating wavelet and RGB losses in Transformer-based models, enhancing high-frequency preservation and achieving superior visual performance on natural images compared to pixel-wise loss training. 

Notably, existing wavelet-based SR methods typically process subbands at a single scale, failing to capture the multi-scale characteristics intrinsic to RSI textures. This fragmented optimization approach separately constrains low-frequency structural components and high-frequency directional details without explicit cross-scale coordination, leading to inconsistent reconstruction of hierarchical spatial patterns. In this paper, we propose a novel Dual-Group Multi-Scale Wavelet Loss to enforce cross-resolution frequency consistency through explicit approximation subband grouping and multi-scale high-frequency aggregation (horizontal/vertical/diagonal).

\section{Methodology} \label{section3}
In this section, we first introduce the overall structure of the proposed GDSR and then describe three important
modules in GDSR, namely RGEG, RDEG, and GDRM. Finally, we detail our designed loss function.

\subsection{Overview of GDSR}

As illustrated in Fig.~\ref{fig_Network}, GDSR begins by extracting the shallow feature $\bm{\mathit{F}}_\text{S}\in\mathbb{R}^{H\times W\times C}$ from the LR input using a convolutional layer. It then derives the deep features $\bm{\mathit{F}}_\text{D\_g}\in\mathbb{R}^{H\times W\times C}$ and $\bm{\mathit{F}}_\text{D\_d}\in\mathbb{R}^{H\times W\times C}$ through parallel branches: global extraction branch and details extraction branch, respectively. These features are subsequently fused using the GDRM. A global residual connection is then applied to combine the low-level and deep features, and finally, the pixel-shuffle method~\cite{shi2016real} is employed for upsampling to generate the SR output.
\begin{figure}[!t]
\centering
\includegraphics[width=3.3in]{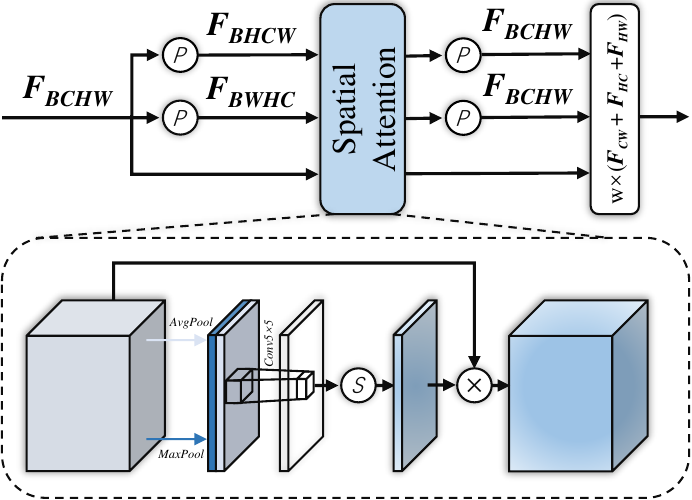}
\captionsetup{font={scriptsize}}
\caption{Structure of PSAM.}
\label{fig_PSAM}
\end{figure}
The SR process of GDSR can be mathematically described as:
\begin{equation}
\bm{\mathit{I}}_{\text{SR}}=\mathrm{Up}(\mathrm{GDRM}(\bm{\mathit{F}}_\text{D\_g},\bm{\mathit{F}}_\text{D\_d})+\bm{\mathit{F}}_\text{S}),
\end{equation}
where $\mathrm{Up}(\cdot)$ denotes the upsampling function and $\mathrm{GDRM}(\cdot)$ denotes the global-detail reconstruction operation. $\bm{\mathit{I}}_{\text{SR}}\in\mathbb{R}^{(rH)\times (rW)\times 3}$ is the reconstructed SR image with a scaling factor of $r$.

\subsection{Residual Details Extraction Group}

Deep convolutional networks have achieved remarkable advancements in SR tasks, with residual structures proving effective in addressing network degradation issues~\cite{edsr}. To harness these advantages, we propose RDEGs, which are structured as sequences of residual units, each referred to as a Residual Convolution Block (RCB). RCBs are specifically designed to enhance the recovery of fine details from LR inputs while maintaining stable and efficient model training.

A single RCB can be represented as:
\begin{equation}
\bm{\mathit{F}}_\text{l+1}=\alpha\bm{\mathit{F}}_\text{l}+\mathrm{W}_\text{1}(\text{LReLU}(\mathrm{W}_\text{0}\bm{\mathit{F}}_\text{l})),
\end{equation}
where $\bm{\mathit{F}}_\text{l}$ and $\bm{\mathit{F}}_\text{l+1}$ represent the input and output feature maps of the $\text{l}$-th RCB, respectively. $\alpha$ denotes a learnable scale factor that controls the amount of information passed through the residual connection. $\mathrm{W}_\text{0}$ and $\mathrm{W}_\text{1}$ denote the convolution operations in the RCB and $\text{LReLU}$ refers to the Leaky ReLU activation function.

We concatenate $\mathrm{N}$ RCBs to form an RDEG module. The RDEG can be mathematically expressed as:
\begin{equation}
\mathrm{RDEG}^\mathrm{k}=\alpha\bm{\mathit{F}}_\text{D\_d}^\text{N-1}+\mathrm{W}_\text{1}(\text{LReLU}(\mathrm{W}_\text{0}(\bm{\mathit{F}}_\text{D\_d}^\text{N-1}))),\mathrm{N}\geq1,
\end{equation}
where $\bm{\mathit{F}}_\text{D\_d}^\text{0}=\mathrm{RDEG}^\text{k-1}$. Here, $\mathrm{RDEG}^\mathrm{k}$  represents the feature map of the $\text{k}$-th RDEG,
$\bm{\mathit{F}}_\text{D\_d}^\text{N-1}$ denotes the output of the $\text{(N-1)}$-th RCB within the $\text{k}$-th RDEG.

\subsection{Residual Global Extraction Group}

\begin{figure}[!t]
\centering
\includegraphics[width=3.4in]{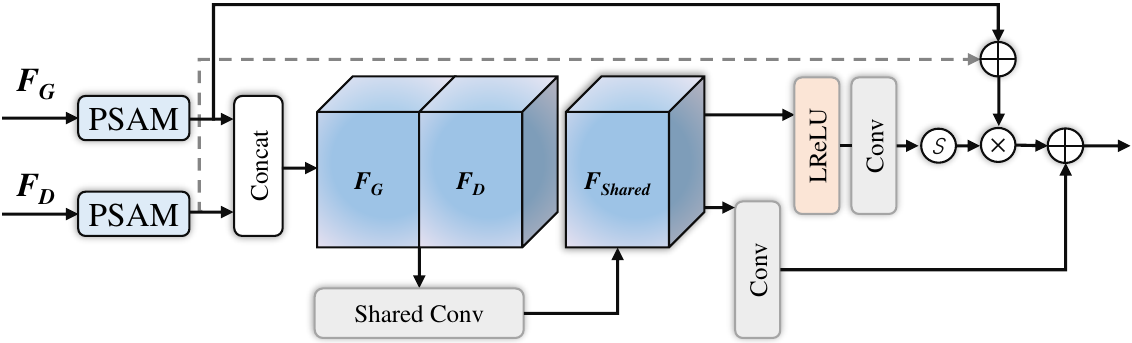}
\captionsetup{font={scriptsize}}
\caption{Structure of GDRM.}
\label{fig_GDRM}
\end{figure}

\begin{figure*}[!t]
\centering
\includegraphics[width=6.8in]{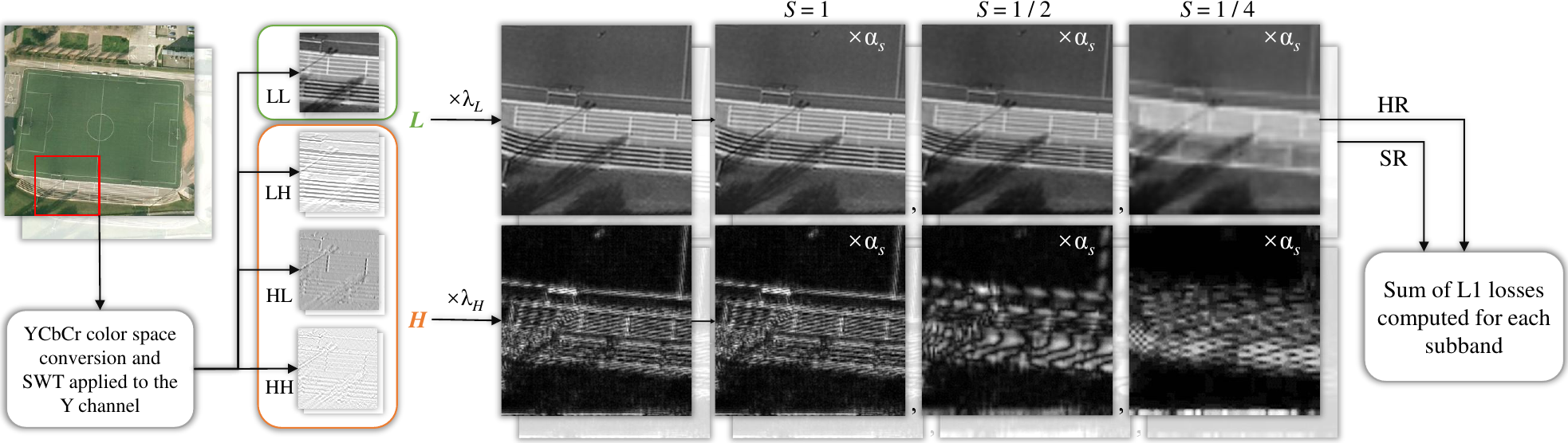}%
\captionsetup{font={scriptsize}}
\caption{Overview of the proposed Dual-Group Multi-Scale Wavelet Loss. Zoom in with brightness-optimized high-frequency subbands for better observation.}
\label{fig_Wavelet_loss}
\end{figure*}

Inspired by the recent success of RWKV in computer vision applications~\cite{duan2025visionrwkv, restorerwkv}, we seek to utilize its strengths in modeling long-range dependencies in RSIs. As illustrated in Fig.~\ref{fig_Network}, the proposed RGEG consists of multiple Residual RWKV Blocks (R-RWKVBs). The spatial mix (sm) module is designed to establish long-range dependencies among tokens across the spatial dimension. Given an input feature flattened into a one-dimensional sequence $\bm{\mathit{F}_\text{in}}\in\mathbb{R}^{T\times C}$, where $T=H\times W$ represents the total number of tokens, the spatial mix module begins by applying layer normalization (LN) followed by the Omni-Shift operation~\cite{restorerwkv}. The output is then passed through three parallel linear projection layers to obtain the receptance $R_s\in\mathbb{R}^{T\times C}$, key $K_s\in\mathbb{R}^{T\times C}$, and value $V_s\in\mathbb{R}^{T\times C}$:
\begin{equation}
\bm{\mathit{F}}_{s}=\text{Omni-Shift}(\mathrm{LN}(\bm{\mathit{F}_\text{in}})),
\end{equation}
\begin{equation}
R_{s}=\bm{\mathit{F}}_{s}W_{R_{s}},\quad K_{s}=\bm{\mathit{F}}_{s}W_{K_{s}},\quad V_{s}=\bm{\mathit{F}}_{s}W_{V_{s}},
\end{equation}
where $W_{R_s}$, $W_{K_s}$, and $W_{V_s}$ are linear projection matrices. The key $K_{s}$ and value $V_{s}$
are then processed through the Re-WKV attention mechanism~\cite{restorerwkv} to compute the global attention output $O_\text{attn}\in\mathbb{R}^{T\times C}$. Finally, the receptance $R_{s}$, after gating through a sigmoid function $\sigma(\cdot)$, modulates the attention output via element-wise multiplication:
\begin{equation}
O_\text{attn}=\text{Re-WKV}(K_s,V_s),
\end{equation}
\begin{equation}
\bm{\mathit{F}}_\text{sm}=(\sigma(R_s)\odot O_\text{attn})W_\text{sm},
\end{equation}
where $\bm{\mathit{F}}_\text{sm}$ represents the spatial mix output, 
$\odot$ denotes element-wise multiplication, and 
$W_\text{sm}$ is the output projection matrix. The channel mix (cm) module focuses on channel-wise feature fusion. Similar to the spatial mix module, it starts with LN and Omni-Shift, and the receptance $R_c\in\mathbb{R}^{T\times C}$, key $K_c\in\mathbb{R}^{T\times C}$, and value $V_c\in\mathbb{R}^{T\times C}$ are computed as follows:
\begin{equation}
\bm{\mathit{F}}_{c}=\text{Omni-Shift}(\mathrm{LN}(\alpha_1\bm{\mathit{F}}_{in}+\bm{\mathit{F}}_\text{sm})),
\end{equation}
\begin{equation}
R_c=\bm{\mathit{F}}_cW_{R_c},\quad K_c=\bm{\mathit{F}}_cW_{K_c},\quad V_c=\gamma(K_c)W_{V_c},
\end{equation}
where $W_{R_c}$, $W_{K_c}$, and $W_{V_c}$ are linear projection matrices, and $\gamma(\cdot)$ is the squared ReLU activation function, which enhances nonlinearity. Notably, the transformation from $\bm{\mathit{F}}_{c}$ to $K_{c}$ to $V_{c}$ involves a multi-layer perception consisting of $W_{K_c}$, $\gamma(\cdot)$, and $W_{V_c}$, facilitating channel-wise feature fusion. The final channel mix output 
$\bm{\mathit{F}}_\text{cm}$ is obtained through:
\begin{equation}
\bm{\mathit{F}}_\text{cm}=(\sigma(R_c)\odot V_c)W_\text{cm},
\end{equation}
where $W_\text{cm}$ is the output projection matrix. Finally, RGEG can be formulated as 
\begin{equation}
R=\alpha_2(\alpha_1\bm{\mathit{F}}_{in}+\bm{\mathit{F}}_\text{sm})+\bm{\mathit{F}}_\text{cm},
\end{equation}
\begin{equation}
\mathrm{RGEG}^{k}=\mathrm{W}R^{l}(\ldots R^{1}(\mathrm{RGEG}^{k-1})\ldots)+\mathrm{RGEG}^{k-1},
\end{equation}
where $\mathrm{RGEG}^{k}$
and $\mathrm{RGEG}^{k-1}$ denote the feature map of the $\mathrm{k}$-th RGEG and $\mathrm{(k\!-\!1)}$-th RGEG, respectively. $\mathrm{R}^{1}$ is the $\mathrm{l}$-th R-RWKVB. $\mathrm{W}$ is a convolutional layer that serves to enhance the translational equivariance of the RWKV layer, while $\alpha_1$ and $\alpha_2$ are learnable scale factors.

\subsection{Global–Detail Reconstruction Module}

Remote sensing images often exhibit intricate spatial details and large-scale contextual information, necessitating a specialized module to harmonize these features for enhanced reconstruction performance. To effectively integrate the complementary features extracted by the RDEG and the RGEG, the GDRM is designed as a fusion mechanism tailored for RSI-SR tasks.

To harmonize and align these two feature representations, the GDRM employs a Permuted Spatial Attention Module (PSAM), which is designed to be computationally efficient and structurally simple. As illustrated in Fig.~\ref{fig_PSAM}, the module begins by applying a permute operation to the input feature map, reorganizing it into three distinct formats, each capturing specific spatial relationships: Height-Width (HW), Channel-Width (CW), and Height-Channel (HC). Spatial Attention~\cite{woo2018cbam} is subsequently applied to each permuted representation independently, enabling the modeling of dependencies along these dimensions. The processed features are then permuted back to their original format, summed together, and scaled by a coefficient. Formally, the aligned feature maps are computed as:
\begin{align}
\bm{\mathit{F}}_\text{CW}, \bm{\mathit{F}}_\text{HC}=\mathrm{Permute}(&\mathrm{SA}(\mathrm{Permute}(\bm{\mathit{F}}_\text{in}))), \notag\\
\bm{\mathit{F}}_\text{HW}=&\mathrm{SA}(\bm{\mathit{F}}_\text{in}),
\end{align}
\begin{equation}
\bm{\mathit{F}}_\text{aligned}=W(\bm{\mathit{F}}_\text{CW}+\bm{\mathit{F}}_\text{HC}+\bm{\mathit{F}}_\text{HW}),
\end{equation}
where $\mathrm{SA}(\cdot)$ denotes the spatial attention operation and $W$ is a fixed coefficient. As illustrated in Fig.~\ref{fig_GDRM}, the GDRM receives two input feature maps: one capturing local details $\bm{\mathit{F}}_\text{D}\in\mathbb{R}^{H\times W\times C}$ from the RDEG branch and another representing global contextual information $\bm{\mathit{F}}_\text{G}\in\mathbb{R}^{H\times W\times C}$ from the RGEG branch. The GDRM then processes these feature maps through a series of operations, which can be expressed as:
\begin{equation}
\bm{\mathit{F}}_\text{G}=\mathrm{PSAM}(\bm{\mathit{F}}_\text{G}), \bm{\mathit{F}}_\text{D}=\mathrm{PSAM}(\bm{\mathit{F}}_\text{D}),
\end{equation}
\begin{align}
\bm{\mathit{F}}_\text{shared}&=\mathrm{W}_\text{shared}(\mathrm{cat}(\bm{\mathit{F}}_\text{G}, \bm{\mathit{F}}_\text{D})), \notag\\
w&=\sigma\mathrm{W}_0(\text{LReLU}(\bm{\mathit{F}}_\text{shared})), \notag\\
b&=\mathrm{W}_1(\bm{\mathit{F}}_\text{shared}),\label{eq16}
\end{align}
\begin{equation}
\bm{\mathit{F}}_\text{GDRM}=w(\bm{\mathit{F}}_\text{G}+\bm{\mathit{F}}_\text{D})+b.\label{eq17}
\end{equation}

\subsection{Dual-Group Multi-Scale Wavelet Loss}

The proposed wavelet loss construction follows the paradigm established in prior studies~\cite{WGSR, Waveletloss}. As illustrated in Fig.~\ref{fig_Wavelet_loss}, initially, both the HR image and SR image are transformed into the YCbCr color space. The stationary wavelet transform (SWT) is subsequently applied exclusively to the luminance (Y) channel, as human visual perception demonstrates higher sensitivity to structural details in this component. Through SWT decomposition, one low-frequency (LF) subband (LL) and three high-frequency (HF) subbands (LH, HL, HH) are generated. The LL subband encapsulates coarse structural information, while the HF components encode directional details: LH captures horizontal features, HL preserves vertical characteristics, and HH retains diagonal patterns. These HF subbands are concatenated along the channel dimension to form a composite HF representation $\bm{\mathit{H}}\in\mathbb{R}^{H\times W\times 3}$, whereas the LL subband is retained as $\bm{\mathit{L}}\in\mathbb{R}^{H\times W\times 1}$.

The wavelet loss function is formulated as the $\mathcal{L}_{1}$ distance between the corresponding subbands of the generated SR image and the HR image. The total wavelet loss is expressed as:
\begin{equation}
\mathcal{L}_{\mathrm{wav}}=\lambda_L\sum_s\alpha_s\|\bm{\mathit{L}}_{\mathrm{SR}}^s-\bm{\mathit{L}}_{\mathrm{HR}}^s\|_1+\lambda_H\sum_s\alpha_s\|\bm{\mathit{H}}_{\mathrm{SR}}^s-\bm{\mathit{H}}_{\mathrm{HR}}^s\|_1,
\end{equation}
where $s\in\{1,\frac{1}{2},\frac{1}{4}\}$ denotes three spatial resolutions, $\bm{\mathit{L}}_{\mathrm{SR}}^s$, $\bm{\mathit{L}}_{\mathrm{HR}}^s$ and $\bm{\mathit{H}}_{\mathrm{SR}}^s$, $\bm{\mathit{H}}_{\mathrm{HR}}^s$ represent the LF/HF subbands of the SR/HR images at scale $s$, $\alpha_s$ are resolution-specific weights satisfying $\sum\alpha_{s}=1$ (Specific settings for $\alpha_s$ can be found in Table~\ref{wavelet_exp}, but generally, arbitrary $\alpha_s$ can be used), and $\lambda_L$, $\lambda_H$ control the global contributions of LF/HF components. The overall loss for the training with wavelet loss is given by
\begin{equation}
\mathcal{L}_{\mathrm{rec}}=\parallel \bm{\mathit{I}}_{\mathrm{HR}}-\bm{\mathit{I}}_{\mathrm{SR}}\parallel_1,
\end{equation}
\begin{equation}
\mathcal{L}=\mathcal{L}_{\mathrm{rec}}+\mathcal{L}_{\mathrm{wav}}.
\end{equation}

\section{Experiment} \label{section4}

\subsection{Datasets and Evaluation}
This paper reports the performance of SR on three RSI datasets, including the publicly available AID~\cite{aid}, UCMerced~\cite{UCMerced}, and the custom RSSRD-QH dataset. 

The AID dataset consists of 30 categories, with the data in each category randomly divided into two equal parts for training and testing. Additionally, 20\% of the training data was reserved as a validation set, resulting in 4,000 training images, 1,000 validation images, and 5,000 test images. Similarly, the UCMerced dataset, comprising 21 categories with a total of 2,100 images, was processed identically, yielding 945 training images, 105 validation images (10\% of the training data), and 1,050 test images.

The RSSRD-QH dataset is located in representative watershed units within the Sanjiangyuan region and the Qinghai Lake surrounding area. The geographical distribution of the counties covered by the dataset, including their longitude and latitude ranges, is summarized in Table~\ref{geographical_info}. The dataset was collected using unmanned aerial vehicles, with flight altitudes carefully controlled between 30 and 50 meters. The average spatial resolution of the captured imagery is 0.01 meters, providing comprehensive coverage of various landscape elements across different elevation gradients, including terraces, floodplains, gentle slopes, and steep slopes within these watershed units. From each Watershed, we randomly selected 50\% of the images for the training set and 50\% for the test set, with 20\% of the training set randomly selected for validation. This resulted in a training set of 2,242 images, a validation set of 561 images, and a test set of 2,804 images. The HR images in the AID and RSSRD-QH datasets were 600 × 600 pixels, while those in the UCMerced dataset were 256 × 256 pixels.

The proposed method, along with other competing methods, was evaluated on the test datasets using two classic full-reference metrics: Peak Signal-to-Noise Ratio (PSNR) and Structural Similarity Index (SSIM)~\cite{ssim}. Additionally, the LPIPS~\cite{lpips} metric was used to capture the perceptual quality of the images. Notably, PSNR and SSIM were calculated based on the luminance channel (Y) of the YCbCr color space, while LPIPS evaluated the perceptual similarity between images by extracting features from a pre-trained deep network, specifically the AlexNet~\cite{krizhevsky2012imagenet}, which was employed in this study.

\subsection{Degradation Model and Implementation Details}

We employed both bicubic downsampling and a comprehensive degradation model (CDM) to simulate LR images during training. Based on existing blind SR methods~\cite{blind1,blind2} and the characteristics of remote sensing sensors~\cite{blind3,refdiff}, the CDM for LR image synthesis included isotropic Gaussian blur, motion blur, scaling with different interpolation methods, additive Gaussian noise, and JPEG compression noise, all applied in random order.

Our GDSR performed deep feature exploration through 4 RGEGs, RDEGs, and GDRMs. Each RGEG consisted of 6 R-RWKVBs, and each RDEG contains 12 RCBs. Empirically, we set the internal channel dimension to $c$ = 96. For the GDSR\_TC and GDSR\_MC, the RGEG branches correspond to the RTSB and ResidualGroup of SwinIR~\cite{swinir} and MambaIR~\cite{mambair}, respectively, with all other settings remaining consistent with GDSR.

During training, 64 × 64 patches were randomly cropped from LR RSIs, along with their HR counterparts scaled by the target scale factor (×2, ×3, ×4). The batch size was set to 16. The initial learning rate was set to $1\times10^{-4}$ and optimized using the Adam optimizer, where $\beta1$ = 0.9 and $\beta2$ = 0.99, with a step learning rate scheduler halving the rate at the 100th epoch over a total of 200 training epochs. All SR models were implemented in the PyTorch framework. The FLOPs results were calculated using an input tensor of size 1 × 3 × 160 × 160, and the computations were performed using torch-operation-counter~\footnote{\url{https://github.com/
SamirMoustafa/torch-operation-counter}}. The inference times were tested on 300 random images with a size of 3 × 160 × 160. All experiments were conducted on an NVIDIA A800 80GB GPU.

\begin{table}[!t]
  \centering
  \captionsetup{font={scriptsize}, labelsep = newline, justification=centering}
  \caption{Geographical Distribution of Counties in the RSSRD-QH Dataset.}
    \begin{tabular}{cccccc}
    \toprule[1.5pt]
    County & Longitude (°E) & Latitude (°N) \\
    \midrule    
    \midrule
    Maqin           & 100.219 to 100.285        & 34.264 to 34.7             \\
    Gande           & 100.047 to 100.719        & 34.054 to 34.204           \\
    Dari            & 98.7845 to 98.996         & 33.485 to 33.925           \\
    Maduo           & 97.929 to 98.129          & 34.668 to 34.698           \\
    Henan           & 101.374 to 101.759        & 34.307 to 34.707           \\
    Gangcha         & 100.406 to 100.509        & 37.351 to 37.671           \\
    Tianjun         & 98.409 to 98.624          & 37.45 to 37.4855           \\
    \bottomrule[1.5pt]
    \end{tabular}%
  \label{geographical_info}%
\end{table}%

\subsection{Ablation Study} \label{sectionAb}

In this section, we discuss the proposed GDSR in depth by investigating the effect of its major components and their variants. All experiments were conducted under the ×3 super-resolution configuration with the $\mathcal{L}_{\mathrm{rec}}$ loss function in the AID~\cite{aid} dataset.

\subsubsection{Effect of Key Modules} \textbf{(a) Effect of RDEG and RGEG.} Ablation studies were conducted to evaluate the impact of varying RCBs in RDEG and R-RWKVBs in RGEG on model performance. As shown in Tables~\ref{num-RCB-R-RWKVB} and~\ref{GDRM}, the baseline model using only the RGEG branch yielded a PSNR of 28.94 dB and SSIM of 0.7652. Introducing a single RCB to RDEG improved performance to 28.96 dB PSNR and 0.7658 SSIM, validating the RCB’s ability to enhance reconstruction quality. Scaling the RCB count to 12 further increased PSNR to 29.03 dB and SSIM to 0.7686, with a computational footprint of 13.17M parameters and 338.73G FLOPs. However, beyond 12 RCBs, performance plateaued (e.g., 29.01 dB with 16 RCBs) or exhibited instability (e.g., 29.03 dB with 20 RCBs), while computational costs grew linearly, reducing FPS from 14.8 to 12.6. This suggests diminishing returns from excessive RCB stacking, necessitating a balance between accuracy and efficiency.

Similarly, experiments on RGEG’s R-RWKVBs revealed analogous trends. Fixing RCBs at 12 and increasing R-RWKVBs from 1 to 8 marginally improved PSNR and SSIM but incurred significant computational overhead, with parameters rising from 10.62M to 14.19M and FPS dropping sharply from 40.9 to 10.5. Notably, the joint configuration of 12 RCBs and 6 R-RWKVBs achieved optimal trade-offs, outperforming isolated branch evaluations (as shown in Table~\ref{GDRM}). These results underscore the complementary roles of RDEG and RGEG in feature refinement while emphasizing the necessity of constrained component counts to mitigate redundancy and computational bloat in practical deployments.

\begin{figure}[!t]
\centering
\includegraphics[width=3.4in]{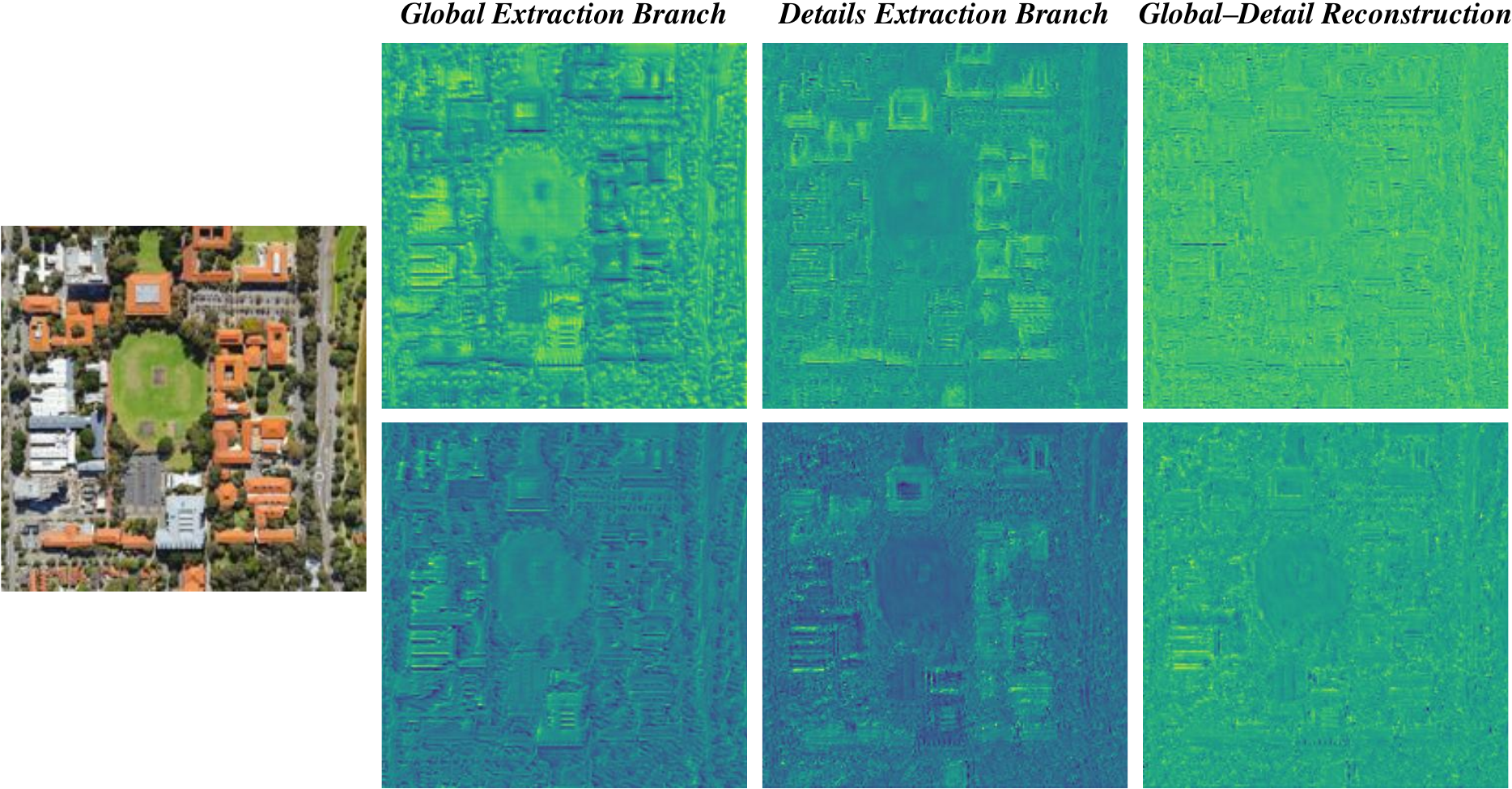}
\captionsetup{font={scriptsize}}
\caption{Feature Visualization Comparisons. The first row shows channel-0 feature maps from the PSAM-processed final outputs of RGEG and RDEG, with the second row depicting channel-1 feature maps from the same modules. Visualizations of GDRM outputs at corresponding network stages are provided in parallel for comparison.}
\label{fig_GDRM_f}
\end{figure}

\begin{table}[!t]
  \centering
  \captionsetup{font={scriptsize}, labelsep = newline, justification=centering}
  \caption{Ablation studies for the number of RCB and R-RWKVB. The configuration with 12 RCBs and 6 R-RWKVBs denotes the proposed GDSR model. Bold indicates the model proposed in this work.}
  \setlength{\tabcolsep}{1.8mm}{
    \begin{tabular}{c|cccccc}
    \toprule[1.5pt]
    RCBs & R-RWKVBs & PSNR $\uparrow$ & SSIM $\uparrow$  & \#Param. & FLOPs & FPS\\
    \midrule    
    \midrule
    1   & \multirow{6}{*}{\makecell{\textbf{6}}}  &  28.96     &  0.7658    &  5.86M     & 151.23G  & 15.7 \\
    4   &   &  28.96     &  0.7657    &  7.85M     & 202.37G  & 15.2 \\
    8   &   &  29.00     &  0.7677    &  10.51M     & 270.55G  & 14.5 \\
    \textbf{12}   &   &  \textbf{29.03}     &  \textbf{0.7686}    &  \textbf{13.17M}     & \textbf{338.73G}  & \textbf{14.8} \\
    16   &   &  29.01     &  0.7680   &  15.83M     & 406.92G  & 13.3 \\
    20   &   &  29.03     &  0.7688   &  18.48M     & 475.10G  & 12.6 \\
    \midrule
    \multirow{4}{*}{\makecell{12}}   &  1 &  29.00     &  0.7676   &  10.62M     & 272.97G  & 40.9 \\
    &  2 &  29.01     &  0.7678   &  11.13M     & 286.12G  & 29.4 \\
    &  4 &  29.01     &  0.7679   &  12.15M     & 312.43G  & 16.2 \\
    &  8 &  29.01     &  0.7679   &  14.19M     & 365.04G  & 10.5 \\
    \bottomrule[1.5pt]
    \end{tabular}%
    }
  \label{num-RCB-R-RWKVB}%
\end{table}%

\textbf{(b) Effect of GDRM.} Ablation experiments in Table~\ref{GDRM} demonstrated that the dual-branch GDSR model outperformed single-branch alternatives, achieving superior PSNR, SSIM, and LPIPS values. This performance gain was attributed to the GDRM, which systematically integrated dual-branch features through its PSAM and adaptive fusion mechanism (Eqs.~\ref{eq16} and~\ref{eq17}). From a spatial-frequency duality perspective, RGEG’s global attention mechanisms implicitly exhibited low-pass filtering characteristics to maintain structural continuity, while RDEG’s convolutional operations inherently emphasized high-frequency components for local detail enhancement~\cite{park2022how}. As visualized in Fig.~\ref{fig_GDRM_f}, the dual-branch structure of GDSR demonstrated complementary feature specialization with GDRM dynamically fusing both to enable efficient reconstruction.

\begin{figure}[!t]
\centering
\includegraphics[width=3.4in]{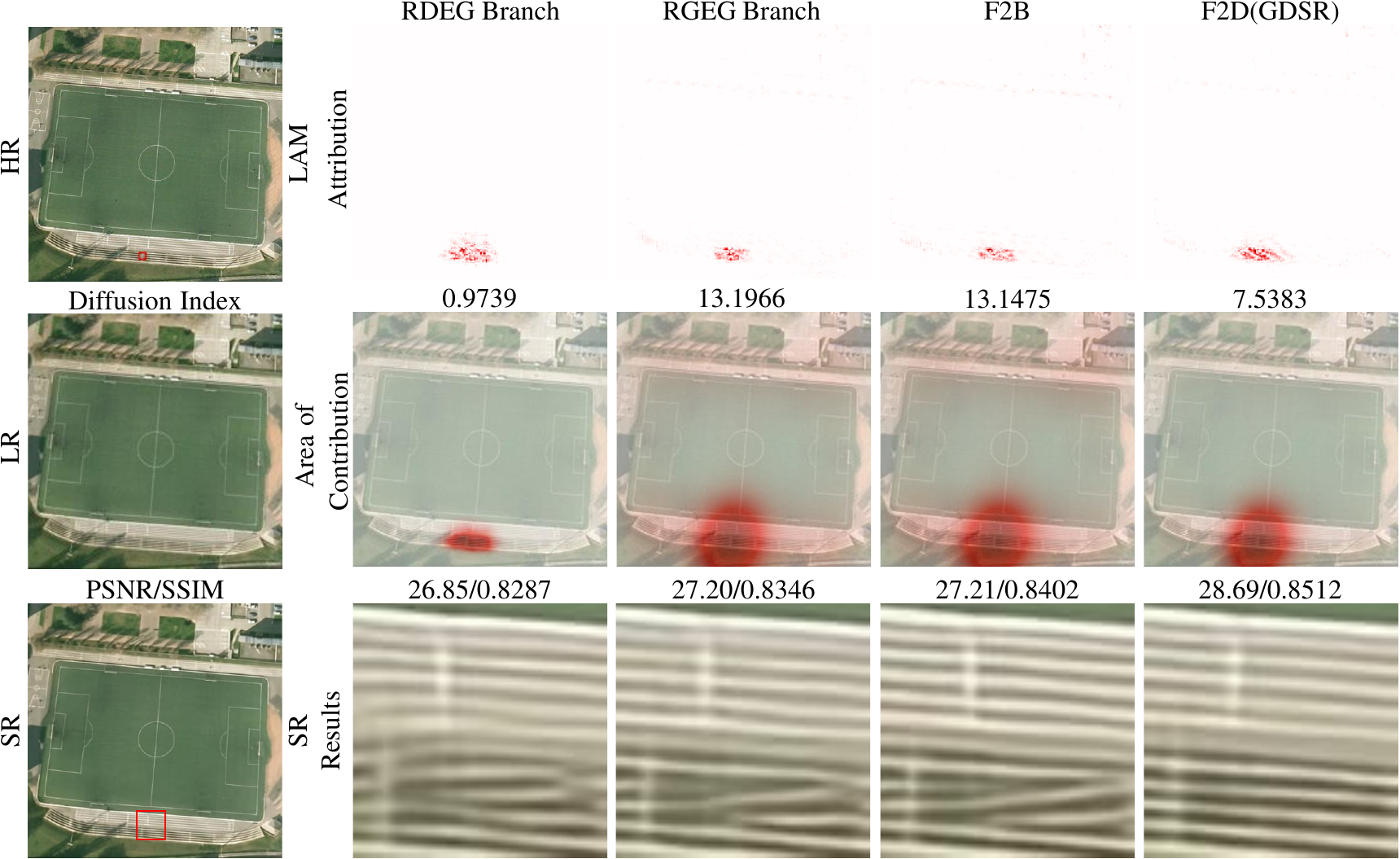}
\captionsetup{font={scriptsize}}
\caption{The visualization of Local Attribution Maps (LAM)~\cite{lam}. The LAM maps represent the importance of each pixel in the input LR image w.r.t. the SR of the patch marked with a red box. The Diffusion Index (DI)~\cite{lam} reflects the range of involved pixels. A higher DI represents a wider range of attention.}
\label{fig_GDRM_LAM}
\end{figure}

Visualizations in Fig.~\ref{fig_GDRM_LAM} revealed that RGEG activated distributed pixels across broader regions, whereas RDEG focused on local patterns. The F2D strategy further enhanced this synergy by routing refined features exclusively to RDEG, prioritizing localized refinement over redundant global propagation. This design addressed a critical theoretical limitation: while RGEG aggregated broader pixel information, effective reconstruction required targeted utilization of spatial relationships within receptive fields. F2D's spectral separation strategy specifically resolved conflicts observed in F2B, where dual-path feedback introduced frequency interference, as global attention diluted high-frequency details while convolutions struggled to recover them.

The quantitative results confirmed that GDSR’s dual-branch framework, coupled with GDRM’s reconstruction, systematically unified heterogeneous features to achieve both perceptual and pixel-wise accuracy. This validated the theoretical rationale that explicitly decoupling global (low-frequency) and local (high-frequency) processing paths, followed by attention-driven fusion, was essential for overcoming the limitations of conventional single-path architectures.

\begin{table}[!t]
  \centering
  \captionsetup{font={scriptsize}, labelsep = newline, justification=centering}
  \caption{Ablation studies for the GDRM. F2B (Feedback to Both RDEG and RGEG) refers to the strategy where the output of the GDRM is sent back to both the RDEG and the RGEG, enabling updates to the RGEG features. F2D (Feedback to RDEG Only) refers to the strategy where the output of the GDRM is sent back only to the RDEG, while the RGEG features continue to propagate unchanged from previous layers. Bold indicates the model proposed in this work.}
    \begin{tabular}{cccc}
    \toprule[1.5pt]
    Method & PSNR (dB) $\uparrow$ & SSIM $\uparrow$  & LPIPS $\downarrow$\\
    \midrule    
    \midrule
    RDEG Branch     &  28.94     &  0.7656    &  0.3554\\
    RGEG Branch     &  28.94     &  0.7652    &  0.3519\\
    w/o PSAM     &  29.01     &  0.7677   &  0.3485\\
    F2B     &  29.01     &  0.7681    &  0.3476\\
    \textbf{F2D(GDSR)}     &  \textbf{29.03}     &  \textbf{0.7686}    &  \textbf{0.3476}\\
    \bottomrule[1.5pt]
    \end{tabular}%
  \label{GDRM}%
\end{table}%

\textbf{(c) Effect of RWKV.} To investigate the effect of RWKV, ablation experiments were conducted by replacing RWKV with Transformer~\cite{swinir} (GDSR\_TC) and Mamba~\cite{mambair} (GDSR\_MC) while keeping other components unchanged. As shown in Table~\ref{effectofrwkv}, the RWKV-based GDSR achieves the best performance in terms of PSNR, SSIM, and LPIPS, outperforming both Transformer and Mamba variants. Notably, RWKV demonstrates superior computational efficiency with 14.8 FPS, significantly exceeding GDSR\_TC and GDSR\_MC. While maintaining moderate parameter size and FLOPs, RWKV achieves an optimal balance between reconstruction quality and computational cost. This validates RWKV’s effectiveness in capturing long-range dependencies while mitigating the quadratic complexity limitations of conventional attention mechanisms.

\textbf{(d) Effect of Network Depth on Performance.} To investigate the effect of network depth on the performance of GDSR, we varied the shared depth of both the RDEG and RGEG branches while keeping other design parameters constant. As illustrated in Table~\ref{depth}, the experimental outcomes consistently demonstrated an enhancement in performance with deeper networks. Moreover, an evaluation of the trade-offs in terms of computational complexity was conducted. As the network depth increased, the model's parameter count, FLOPS, and FPS were also scaled accordingly. These results supported the effectiveness of GDSR's dual-branch design, where the shared depth allowed for both high-quality super-resolution and manageable computational overhead.

\begin{table}[!t]
  \centering
  \captionsetup{font={scriptsize}, labelsep = newline, justification=centering}
  \caption{Ablation studies for the RWKV.}
    \begin{tabular}{cccccc}
    \toprule[1.5pt]
    Method & PSNR/SSIM $\uparrow$ & LPIPS $\downarrow$  & \#Param. & FLOPs & FPS\\
    \midrule    
    \midrule
    GDSR\_TC     &  28.89/0.7633     &  0.3565    &  12.83M     & 337.26G  & 7.6 \\
    GDSR\_MC     &  28.97/0.7663     &  0.3514    &  13.99M     & 350.35G  & 9.1 \\
    \textbf{GDSR}     &  \textbf{29.03/0.7686}     &  \textbf{0.3476}    &  \textbf{13.17M}     & \textbf{338.73G}  & \textbf{14.8} \\
    \bottomrule[1.5pt]
    \end{tabular}%
  \label{effectofrwkv}%
\end{table}%

\begin{table}[!t]
  \centering
  \captionsetup{font={scriptsize}, labelsep = newline, justification=centering}
  \caption{Ablation studies for network depth on performance. Bold indicates the model proposed in this work.}
    \begin{tabular}{cccccc}
    \toprule[1.5pt]
    Depth & PSNR/SSIM $\uparrow$ & LPIPS $\downarrow$  & \#Param. & FLOPs & FPS\\
    \midrule    
    \midrule
    1     &  28.89/0.7633     &  0.3565    &  3.65M     & 94.06G  & 53.0 \\
    2     &  28.97/0.7663     &  0.3514    &  6.82M     & 175.62G  & 27.3 \\
    \textbf{4}     &  \textbf{29.03/0.7686}     &  \textbf{0.3476}    &  \textbf{13.17M}     & \textbf{338.73G}  & \textbf{14.8} \\
    6     &  29.03/0.7686     &  0.3463    &  19.52M     & 501.85G  & 9.3 \\
    8     &  29.04/0.7692     &  0.3456   &  25.86M     & 664.96G  & 7.0 \\
    10     &  29.05/0.7695     &  0.3450   &  32.21M     & 828.07G  & 5.6 \\
    \bottomrule[1.5pt]
    \end{tabular}%
  \label{depth}%
\end{table}%

\subsubsection{Effect of Wavelet Loss}\label{secwavelet}
\begin{figure}[!t]
\centering
\includegraphics[width=3.4in]{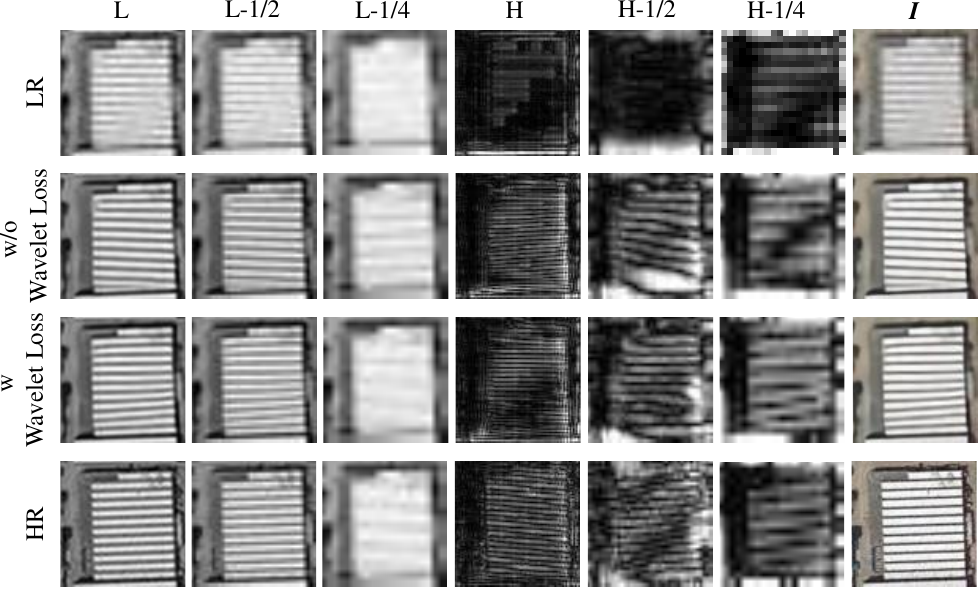}
\captionsetup{font={scriptsize}}
\caption{Visualization of Dual-Group Multi-Scale Wavelet Loss effectiveness in GDSR. Brightness-optimized high-frequency subbands for better observation.}
\label{fig_Wavelet_r}
\end{figure}

Ablation experiments were conducted across multiple configurations to evaluate the effectiveness of incorporating wavelet loss in training. Quantitative results (Table~\ref{wavelet_exp}) demonstrate that our full configuration achieves the highest PSNR and SSIM while attaining the lowest LPIPS, outperforming both baselines and the dual-group-only variant. Visual evidence in Fig.~\ref{fig_Wavelet_r} further corroborates these findings, the inclusion of the proposed Dual-Group Multi-Scale Wavelet Loss significantly improved the reconstruction, particularly in the high-frequency components. This improvement is attributed to the synergistic integration of dual-coefficient grouping strategy and cross-resolution frequency consistency, which collectively overcome the rigidity of fixed-scale wavelet methods that process subbands in isolation. These results highlight Wavelet Loss’s role in guiding the model to recover intricate structures and accurately reconstruct complex textures, thereby enhancing overall image quality.

However, as evidenced in Table~\ref{wavelet_exp}, all wavelet loss variants exhibit higher LPIPS values compared to the $\mathcal{L}_{\mathrm{rec}}$ baseline, indicating an inherent challenge in reconciling frequency-domain fidelity with perceptual authenticity. Specifically, the uniform coefficient configuration~\cite{Waveletloss} yields the worst LPIPS due to its undifferentiated weighting. Our dual-group multi-scale design strategically mitigates this tradeoff. While the proposed method’s LPIPS remains marginally higher than $\mathcal{L}_{\mathrm{rec}}$, it achieves the lowest LPIPS among wavelet-based approaches. The resultant configuration delivers superior PSNR/SSIM with minimal perceptual compromise, demonstrating that while frequency-domain constraints inherently elevate LPIPS, our method optimally balances this tradeoff with substantial PSNR/SSIM gains.

\begin{table}[!t]
  \centering
  \captionsetup{font={scriptsize}, labelsep = newline, justification=centering}
  \caption{Ablation studies for the proposed Dual-Group Multi-Scale Wavelet Loss. We compare our Dual-Group Multi-Scale Wavelet Loss with two baseline configurations from~\cite{Waveletloss}: (1) uniform coefficient allocation (all LL/LH/HL/HH subbands with $\lambda=0.05$) and (2) asymmetric coefficients (LL/HH $\lambda=0.05$, LH/HL $\lambda=0.01$) with isolated subband processing. Proposed loss parameters: $\lambda_L=0.05$, $\lambda_H=0.01$, $\alpha=\{0.6,0.3,0.1\}$ for $s=\{1,\frac{1}{2},\frac{1}{4}\}$. All evaluated wavelet-based methods employ the “sym19” wavelet with 2-level decomposition.}
  \setlength{\tabcolsep}{1.0mm}{
    \begin{tabular}{cccc}
    \toprule[1.5pt]
    Method & PSNR (dB) $\uparrow$ & SSIM $\uparrow$  & LPIPS $\downarrow$\\
    \midrule    
    \midrule
    $\mathcal{L}_{\mathrm{rec}}$ loss     &  29.0329     &  0.7686    &  0.3476\\
    Baseline ~\cite{Waveletloss}: Uniform coefficients     &  29.0182     &  0.7676    &  0.3565\\
    Baseline ~\cite{Waveletloss}: Asymmetric coefficients    &  29.0413     &  0.7687    &  0.3534\\
    Proposed: Dual-Group     &  29.0468     &  0.7690   &  0.3535\\
    \textbf{Proposed: Dual-Group Multi-Scale}     &   \textbf{29.0514}     &   \textbf{0.7694}    &   \textbf{0.3519}\\
    \bottomrule[1.5pt]
    \end{tabular}%
    }
  \label{wavelet_exp}%
\end{table}%

\begin{table*}[!t]
  \centering
  \captionsetup{font={scriptsize}, labelsep = newline, justification=centering}
  \caption{Quantitative comparison on AID~\cite{aid} and UCMerced~\cite{UCMerced} test set in terms of PSNR, SSIM, and LPIPS, where the best performance is highlighted in \textcolor{red}{\textbf{red}}.}
  \renewcommand{\arraystretch}{1.2}
  \setlength{\tabcolsep}{2.5mm}{

    \begin{tabular}{c|c|cccccccccc}
    \bottomrule[1.5pt]
    \multirow{2}{*}{Methods} & \multirow{2}{*}{Type} & \multicolumn{3}{c}{AID x2} & \multicolumn{3}{c}{AID x3} &  \multicolumn{3}{c}{AID x4} &  \\
    
    \cline{3-12}
          &       & PSNR $\uparrow$  & SSIM $\uparrow$  & LPIPS $\downarrow$ & PSNR $\uparrow$  & SSIM $\uparrow$  & LPIPS $\downarrow$ & PSNR $\uparrow$  & SSIM $\uparrow$  & LPIPS $\downarrow$ \\
    \hline
    \hline
    Bicubic & \multirow{13}{*}{\makecell{Bicubic \\ Synthetic} }     & 30.48 & 0.8269 & 0.2769  & 27.89 & 0.7238 & 0.4400
      & 26.43 & 0.6541 & 0.5640 \\
    SRCNN~\cite{srcnn}  &  & 30.87 & 0.8352 & 0.2711
      & 28.22 & 0.7363 & 0.3973      & 26.68 & 0.6679 & 0.4862 \\
    EDSR~\cite{edsr}  &  & 31.70 & 0.8560 & 0.2416
      & 28.94 & 0.7653 & 0.3513      & 27.30 & 0.6994 & 0.4381 \\
    SRGAN~\cite{srgan}  &  & 31.57 & 0.8523 & 0.2481
      & 28.80 & 0.7597 & 0.3601      & 27.20 & 0.6939 & 0.4466 \\
    SwinIR~\cite{swinir}   &  & 31.71 & 0.8567 & 0.2370
      & 28.95 & 0.7658 & 0.3512      & 27.32 & 0.6993 & 0.4373 \\
    HAT~\cite{hat} &  & 31.77 & 0.8579 & 0.2351
     & 28.99 & 0.7671 & 0.3479      & 27.34 & 0.7004 & 0.4342 \\
    HAUNet~\cite{HAUNet} &  & 31.48 & 0.8507 & 0.2503
      & 28.75 & 0.7580 & 0.3657      & 27.15 & 0.6913 & 0.4523 \\
    SPT~\cite{SPT} &  & 31.61 & 0.8547 & 0.2391
      & 28.86 & 0.7620 & 0.3553      & 27.26 & 0.6962 & 0.4418 \\
    ConvFormerSR~\cite{ConvFormerSR} &  & 31.64 & 0.8551 & 0.2298
      & 28.90 & 0.7637 & 0.3516      & 27.26 & 0.6964 & 0.4396 \\
    MambaIR~\cite{mambair}  &  & 31.78 & 0.8583 & 0.2354
      & 29.02 & 0.7684 & 0.3487      & 27.37 & 0.7019 & 0.4348 \\
    FreMamba~\cite{FreMamba} &  & 31.62 & 0.8558 & \textcolor{red}{\textbf{0.2215}}
      & 28.96 & 0.7672 & \textcolor{red}{\textbf{0.3410}}      & 27.32 & 0.7010 & \textcolor{red}{\textbf{0.4255}} \\
    \rowcolor{mygray}\textbf{GDSR (Ours)} &  & \textcolor{red}{\textbf{31.81}} & \textcolor{red}{\textbf{0.8592}} & 0.2366
      & \textcolor{red}{\textbf{29.05}} & \textcolor{red}{\textbf{0.7694}} & 0.3519      & \textcolor{red}{\textbf{27.39}} & \textcolor{red}{\textbf{0.7022}} & 0.4376 \\

    \hline
    
    Bicubic & \multirow{13}{*}{\makecell{CDM \\ Synthetic} }     & 24.11 & 0.5545 & 0.6046  & 23.15 & 0.5238 & 0.6911
      & 22.54 & 0.5091 & 0.7543 \\
    SRCNN~\cite{srcnn}  &  & 24.44 & 0.5854 & 0.5836
      & 23.38 & 0.5455 & 0.6569      & 22.70 & 0.5245 & 0.7100 \\
    EDSR~\cite{edsr}  &  & 24.90 & 0.6153 & 0.5341
      & 23.91 & 0.5698 & 0.6139      & 23.16 & 0.5444 & 0.6753 \\
    SRGAN~\cite{srgan}  &  & 24.78 & 0.6070 & 0.5609
      & 23.78 & 0.5648 & 0.6314      & 23.02 & 0.5396 & 0.6909 \\
    SwinIR~\cite{swinir}   &  & 24.95 & 0.6144 & 0.5353
      & 23.89 & 0.5705 & 0.6169      & 23.17 & 0.5452 & 0.6737 \\
    HAT~\cite{hat} &  & 25.06 & 0.6183 & 0.5276
     & 23.99 & 0.5737 & 0.6080      & 23.26 & 0.5475 & 0.6671 \\
    HAUNet~\cite{HAUNet} &  & 25.03 & 0.6169 & 0.5287
      & 23.90 & 0.5692 & 0.6185      & 23.19 & 0.5441 & 0.6783 \\
    SPT~\cite{SPT} &  & 24.86 & 0.6086 & 0.5457
      & 23.88 & 0.5679 & 0.6257     & 23.20 & 0.5447 & 0.6749 \\
    ConvFormerSR~\cite{ConvFormerSR} &  & 25.04 & 0.6137 & 0.5378
      & 23.97 & 0.5710 & 0.6118      & 23.23 & 0.5448 & 0.6742 \\
    MambaIR~\cite{mambair}  &  & 25.17 & 0.6230 & 0.5120
      & 24.07 & 0.5763 & \textcolor{red}{\textbf{0.6000}}      & 23.32 & 0.5497 & \textcolor{red}{\textbf{0.6590}} \\
    FreMamba~\cite{FreMamba} &  & 25.00 & 0.6227 & \textcolor{red}{\textbf{0.4977}}
      & 23.89 & 0.5692 & 0.6148      & 23.06 & 0.5403 & 0.6735 \\
    \rowcolor{mygray}\textbf{GDSR (Ours)} &  & \textcolor{red}{\textbf{25.28}} & \textcolor{red}{\textbf{0.6256}} & 0.5225
      & \textcolor{red}{\textbf{24.12}} & \textcolor{red}{\textbf{0.5771}} & 0.6106      & \textcolor{red}{\textbf{23.39}} & \textcolor{red}{\textbf{0.5510}} & 0.6664 \\

    \hline
    
    \multicolumn{1}{c}{ } & \multirow{1}{*}{ } & \multicolumn{3}{c}{UCMerced x2} & \multicolumn{3}{c}{UCMerced x3} &  \multicolumn{3}{c}{UCMerced x4} &  \\
    
    \hline
    
    Bicubic & \multirow{13}{*}{\makecell{Bicubic \\ Synthetic} }     & 31.10 & 0.8892 & 0.2013  & 27.59 & 0.7731 & 0.3548
      & 25.73 & 0.6813 & 0.4838 \\
    SRCNN~\cite{srcnn}  &  & 32.33 & 0.9102 & 0.1635
      & 28.48 & 0.8028 & 0.2919      & 26.39 & 0.7137 & 0.3944 \\
    EDSR~\cite{edsr}  &  & 34.31 & 0.9337 & 0.0951
      & 29.98 & 0.8464 & 0.2048      & 27.70 & 0.7644 & 0.3002 \\
    SRGAN~\cite{srgan}  &  & 34.07 & 0.9317 & 0.1007
      & 29.79 & 0.8427 & 0.2207      & 27.53 & 0.7571 & 0.3181 \\
    SwinIR~\cite{swinir}   &  & 34.35 & 0.9344 & 0.0942
      & 30.06 & 0.8514 & 0.2018      & 27.75 & 0.7684 & 0.2956 \\
    HAT~\cite{hat} &  & 34.30 & 0.9339 & 0.0936
     & 30.08 & \textcolor{red}{\textbf{0.8527}} & 0.1979      & 27.77 & \textcolor{red}{\textbf{0.7700}} & 0.2914 \\
    HAUNet~\cite{HAUNet} &  & 33.68 & 0.9270 & 0.1113
      & 29.47 & 0.8343 & 0.2333      & 27.25 & 0.7479 & 0.3381 \\
    SPT~\cite{SPT} &  & 34.12 & 0.9319 & 0.0968
      & 29.79 & 0.8439 & 0.2115      & 27.55 & 0.7613 & 0.3072 \\
    ConvFormerSR~\cite{ConvFormerSR} &  & 34.27 & 0.9330 & 0.0945
      & 29.89 & 0.8461 & 0.2123      & 27.59 & 0.7614 & 0.3109 \\
    MambaIR~\cite{mambair}  &  & 34.52 & \textcolor{red}{\textbf{0.9356}} & \textcolor{red}{\textbf{0.0898}}
      & 30.14 & 0.8513 & \textcolor{red}{\textbf{0.1934}}      & 27.75 & 0.7694 & \textcolor{red}{\textbf{0.2874}} \\
    FreMamba~\cite{FreMamba} &  & 33.87 & 0.9283 & 0.1053
      & 29.95 & 0.8471 & 0.2025      & 27.61 & 0.7639 & 0.2992 \\
    \rowcolor{mygray}\textbf{GDSR (Ours)} &  & \textcolor{red}{\textbf{34.53}} & 0.9354 & 0.0918
      & \textcolor{red}{\textbf{30.15}} & 0.8500 & 0.2057      & \textcolor{red}{\textbf{27.79}} & 0.7683 & 0.3099 \\

    \hline
    
    Bicubic & \multirow{13}{*}{\makecell{CDM \\ Synthetic} }     & 22.46 & 0.5128 & 0.6323  & 21.35 & 0.4678 & 0.7330
      & 20.86 & 0.4558 & 0.7846 \\
    SRCNN~\cite{srcnn}  &  & 22.71 & 0.5392 & 0.6277
      & 21.48 & 0.4860 & 0.7077      & 20.96 & 0.4702 & 0.7593 \\
    EDSR~\cite{edsr}  &  & 22.99 & 0.5635 & 0.5485
      & 21.67 & 0.5056 & 0.6402      & 21.10 & 0.4850 & 0.6917 \\
    SRGAN~\cite{srgan}  &  & 22.96 & 0.5603 & 0.5675
      & 21.65 & 0.5041 & 0.6569      & 21.10 & 0.4836 & 0.7051 \\
    SwinIR~\cite{swinir}   &  & 23.09 & 0.5682 & 0.5496
      & 21.77 & 0.5096 & 0.6481      & 21.18 & 0.4882 & 0.6900 \\
    HAT~\cite{hat} &  & 23.22 & 0.5758 & 0.5338
     & 21.82 & 0.5129 & 0.6374      & 21.18 & 0.4895 & 0.6851 \\
    HAUNet~\cite{HAUNet} &  & 23.03 & 0.5620 & 0.5635
      & 21.69 & 0.5033 & 0.6655      & 21.07 & 0.4824 & 0.7117 \\
    SPT~\cite{SPT} &  & 23.22 & 0.5724 & 0.5464
      & 21.82 & 0.5083 & 0.6565      & 21.26 & 0.4884 & 0.7055 \\
    ConvFormerSR~\cite{ConvFormerSR} &  & 23.30 & 0.5739 & 0.5453
      & 21.91 & 0.5124 & 0.6459      & \textcolor{red}{\textbf{21.27}} & 0.4898 & 0.6951 \\
    MambaIR~\cite{mambair}  &  & 23.23 & 0.5778 & \textcolor{red}{\textbf{0.5254}}
      & 21.76 & 0.5129 & 0.6247      & 21.16 & \textcolor{red}{\textbf{0.4903}} & 0.6749 \\
    FreMamba~\cite{FreMamba} &  & 23.41 & 0.5783 & 0.5287
      & 21.73 & 0.5129 & \textcolor{red}{\textbf{0.6200}}      & 21.12 & 0.4894 & \textcolor{red}{\textbf{0.6723}} \\
    \rowcolor{mygray}\textbf{GDSR (Ours)} &  & \textcolor{red}{\textbf{23.45}} & \textcolor{red}{\textbf{0.5836}} & 0.5304
      & \textcolor{red}{\textbf{21.93}} & \textcolor{red}{\textbf{0.5156}} & 0.6323      & 21.23 & 0.4894 & 0.6828 \\
      
    \toprule[1.5pt]
    \end{tabular}%
    }

  \label{all-res}%
\end{table*}%

\begin{table*}[!t]
  \centering
  \captionsetup{font={scriptsize}, labelsep = newline, justification=centering}
  \caption{Quantitative comparison on RSSRD-QH test set in terms of PSNR, SSIM, and LPIPS, where the best performance is highlighted in \textcolor{red}{\textbf{red}}.}
  \renewcommand{\arraystretch}{1.2}
  \setlength{\tabcolsep}{2.5mm}{

    \begin{tabular}{c|c|cccccccccc}
    \bottomrule[1.5pt]
    \multirow{2}{*}{Methods} & \multirow{2}{*}{Type} & \multicolumn{3}{c}{RSSRD-QH x2} & \multicolumn{3}{c}{RSSRD-QH x3} &  \multicolumn{3}{c}{RSSRD-QH x4} &  \\
    
    \cline{3-12}
          &       & PSNR $\uparrow$  & SSIM $\uparrow$  & LPIPS $\downarrow$ & PSNR $\uparrow$  & SSIM $\uparrow$  & LPIPS $\downarrow$ & PSNR $\uparrow$  & SSIM $\uparrow$  & LPIPS $\downarrow$ \\
    \hline
    \hline
    Bicubic & \multirow{13}{*}{\makecell{Bicubic \\ Synthetic} }     & 29.37 & 0.8149 & \textcolor{red}{\textbf{0.2564}}  & 26.45 & 0.6447 & 0.4243
      & 24.92 & 0.5224 & 0.5764 \\
    SRCNN~\cite{srcnn}  &  & 29.58 & 0.8218 & 0.3028
      & 26.54 & 0.6512 & 0.4241      & 24.98 & 0.5268 & 0.5613 \\
    EDSR~\cite{edsr}  &  & 29.83 & 0.8299 & 0.2838
      & 26.70 & 0.6598 & 0.4141      & 25.10 & 0.5343 & 0.5487 \\
    SRGAN~\cite{srgan}  &  & 29.80 & 0.8293 & 0.2859
      & 26.68 & 0.6586 & 0.4170      & 25.08 & 0.5333 & 0.5522 \\
    SwinIR~\cite{swinir}   &  & 29.83 & 0.8305 & 0.2844
      & 26.70 & 0.6597 & 0.4133      & 25.10 & 0.5349 & 0.5474 \\
    HAT~\cite{hat} &  & 29.84 & 0.8307 & 0.2836
     & 26.71 & 0.6602 & 0.4111      & 25.10 & 0.5352 & 0.5454 \\
    HAUNet~\cite{HAUNet} &  & 29.79 & 0.8289 & 0.2888
      & 26.65 & 0.6569 & 0.4177      & 25.07 & 0.5326 & 0.5494 \\
    SPT~\cite{SPT} &  & 29.83 & 0.8306 & 0.2875
      & 26.69 & 0.6602 & 0.4161      & 25.10 & 0.5349 & 0.5478 \\
    ConvFormerSR~\cite{ConvFormerSR} &  & 29.82 & 0.8310 & 0.2786
      & 26.69 & 0.6603 & \textcolor{red}{\textbf{0.4090}}      & 25.10 & 0.5354 & 0.5452 \\
    MambaIR~\cite{mambair}  &  & 29.86 & \textcolor{red}{\textbf{0.8311}} & 0.2825
      & 26.72 & \textcolor{red}{\textbf{0.6607}} & 0.4124      & 25.11 & 0.5359 & 0.5445 \\
    FreMamba~\cite{FreMamba} &  & 29.80 & 0.8300 & 0.2811
      & 26.70 & 0.6601 & 0.4112      & 25.11 & \textcolor{red}{\textbf{0.5366}} & \textcolor{red}{\textbf{0.5414}} \\
    \rowcolor{mygray}\textbf{GDSR (Ours)} &  & \textcolor{red}{\textbf{29.87}} & 0.8310 & 0.2823
      & \textcolor{red}{\textbf{26.73}} & 0.6601 & 0.4130      & \textcolor{red}{\textbf{25.12}} & 0.5354 & 0.5465 \\

    \hline
    
    Bicubic & \multirow{13}{*}{\makecell{CDM \\ Synthetic} }     & 23.10 & 0.3991 & \textcolor{red}{\textbf{0.5996}}  & 22.34 & 0.3525 & \textcolor{red}{\textbf{0.6993}}
      & 21.85 & 0.3284 & \textcolor{red}{\textbf{0.7764}} \\
    SRCNN~\cite{srcnn}  &  & 23.27 & 0.4159 & 0.6520
      & 22.44 & 0.3588 & 0.7523      & 21.91 & 0.3311 & 0.8078 \\
    EDSR~\cite{edsr}  &  & 23.67 & 0.4485 & 0.6269
      & 22.74 & 0.3817 & 0.7283      & 22.15 & 0.3444 & 0.8028 \\
    SRGAN~\cite{srgan}  &  & 23.57 & 0.4387 & 0.6442
      & 22.64 & 0.3759 & 0.7421      & 22.09 & 0.3408 & 0.8174 \\
    SwinIR~\cite{swinir}   &  & 23.68 & 0.4463 & 0.6232
      & 22.72 & 0.3780 & 0.7432      & 22.15 & 0.3440 & 0.8104 \\
    HAT~\cite{hat} &  & 23.75 & 0.4528 & 0.6120
     & 22.79 & 0.3823 & 0.7312      & 22.18 & 0.3456 & 0.8061 \\
    HAUNet~\cite{HAUNet} &  & 23.72 & 0.4501 & 0.6136
      & 22.76 & 0.3797 & 0.7360      & 22.16 & 0.3442 & 0.8096 \\
    SPT~\cite{SPT} &  & 23.58 & 0.4353 & 0.6512
      & 22.73 & 0.3743 & 0.7520      & 22.14 & 0.3415 & 0.8144 \\
    ConvFormerSR~\cite{ConvFormerSR} &  & 23.67 & 0.4398 & 0.6267
      & 22.77 & 0.3779 & 0.7365      & 22.18 & 0.3436 & 0.8051 \\
    MambaIR~\cite{mambair}  &  & \textcolor{red}{\textbf{23.82}} & \textcolor{red}{\textbf{0.4568}} & 0.6036
      & 22.84 & 0.3835 & 0.7387      & 22.23 & 0.3476 & 0.8003 \\
    FreMamba~\cite{FreMamba} &  & 23.59 & 0.4329 & 0.6009
      & 22.60 & 0.3660 & 0.7242      & 22.06 & 0.3358 & 0.7894 \\
    \rowcolor{mygray}\textbf{GDSR (Ours)} &  & 23.75 & 0.4454 & 0.6197
      & \textcolor{red}{\textbf{22.87}} & \textcolor{red}{\textbf{0.3864}} & 0.7308      & \textcolor{red}{\textbf{22.27}} & \textcolor{red}{\textbf{0.3488}} & 0.8044 \\

    \toprule[1.5pt]
    \end{tabular}%
    }

  \label{qh-res}%
\end{table*}%

\begin{table*}[!t]
  \centering
  \captionsetup{font={scriptsize}, labelsep = newline, justification=centering}
  \caption{Quantitative comparison with SOTA SR methods across 30 scene categories on AID~\cite{aid}, where the best and second best PSNR/SSIM performance are highlighted in \textcolor{red}{\textbf{red}} and \textcolor{blue}{\textbf{blue}}, respectively.}
  \scalebox{0.92}{
    \setlength{\tabcolsep}{1.2mm}{ 
  \begin{tabular}{c|cccccccccccccccc}
    \toprule[1.5pt]
    \multirow{2}[4]{*}{Categoties} & \multicolumn{2}{c}{Bicubic} & \multicolumn{2}{c}{SRGAN~\cite{srgan}} & \multicolumn{2}{c}{EDSR~\cite{edsr}} & \multicolumn{2}{c}{HAT~\cite{hat}} & \multicolumn{2}{c}{MambaIR~\cite{mambair}} & \multicolumn{2}{c}{ConvFormerSR~\cite{ConvFormerSR}} & \multicolumn{2}{c}{FreMamba~\cite{FreMamba}} & \multicolumn{2}{c}{\textbf{GDSR (Ours)}} \\
\cmidrule{2-17}          & PSNR$\uparrow$  & SSIM$\uparrow$  & PSNR$\uparrow$  & SSIM$\uparrow$  & PSNR$\uparrow$  & SSIM$\uparrow$  & PSNR$\uparrow$  & SSIM$\uparrow$  & PSNR$\uparrow$  & SSIM$\uparrow$  & PSNR$\uparrow$  & SSIM$\uparrow$  & PSNR$\uparrow$  & SSIM$\uparrow$ & PSNR$\uparrow$  & SSIM$\uparrow$\\
\midrule
\midrule
    Airport & 27.43  & 0.7387  & 28.48  & 0.7765  & 28.63  & 0.7823  & 28.69  & 0.7837  & \textcolor{blue}{\textbf{28.72}}  & \textcolor{blue}{\textbf{0.7856}}  & 28.59  & 0.7805  & 28.68  & 0.7851  & \textcolor{red}{\textbf{28.74}}  & \textcolor{red}{\textbf{0.7865}}  \\
    Bare Land & 34.45  & 0.8344  & 34.84  & 0.8422  & 34.92  & 0.8439  & 34.94  & 0.8445  & \textcolor{blue}{\textbf{34.96}}  & \textcolor{blue}{\textbf{0.8449}}  & 34.91  & 0.8436  & 34.93  & 0.8444  & \textcolor{red}{\textbf{34.98}}  & \textcolor{red}{\textbf{0.8455}}  \\
    Baseball Field & 29.19  & 0.7786  & 30.27  & 0.8094  & 30.42  & 0.8137  & 30.45  & 0.8144  & \textcolor{blue}{\textbf{30.52}}  & \textcolor{blue}{\textbf{0.8160}}  & 30.35  & 0.8123  & 30.43  & 0.8147  & \textcolor{red}{\textbf{30.55}}  & \textcolor{red}{\textbf{0.8169}}  \\
    Beach & 31.32  & 0.7947  & 31.86  & 0.8082  & 31.94  & 0.8103  & 31.99  & 0.8117  & \textcolor{blue}{\textbf{32.00}}  & \textcolor{blue}{\textbf{0.8118}}  & 31.94  & 0.8096  & 31.95  & 0.8113  & \textcolor{red}{\textbf{32.03}}  & \textcolor{red}{\textbf{0.8124}}  \\
    Bridge & 30.19  & 0.8063  & 31.27  & 0.8312  & 31.46  & 0.8353  & 31.52  & 0.8363  & \textcolor{blue}{\textbf{31.56}}  & \textcolor{blue}{\textbf{0.8372}}  & 31.42  & 0.8339  & 31.52  & 0.8368  & \textcolor{red}{\textbf{31.59}}  & \textcolor{red}{\textbf{0.8378}}  \\
    Center & 25.76  & 0.6980  & 27.18  & 0.7590  & 27.38  & 0.7665  & 27.47  & 0.7694  & \textcolor{blue}{\textbf{27.51}}  & \textcolor{blue}{\textbf{0.7706}}  & 27.33  & 0.7645  & 27.43  & 0.7692  & \textcolor{red}{\textbf{27.52}}  & \textcolor{red}{\textbf{0.7714}}  \\
    Church & 23.09  & 0.6084  & 24.41  & 0.6807  & 24.56  & 0.6895  & 24.61  & 0.6920  & \textcolor{blue}{\textbf{24.65}}  & \textcolor{blue}{\textbf{0.6941}}  & 24.52  & 0.6875  & 24.56  & 0.6919  & \textcolor{red}{\textbf{24.68}}  & \textcolor{red}{\textbf{0.6954}}  \\
    Commercial & 26.72  & 0.7178  & 27.59  & 0.7582  & 27.73  & 0.7645  & 27.77  & 0.7660  & \textcolor{blue}{\textbf{27.81}}  & \textcolor{blue}{\textbf{0.7679}}  & 27.68  & 0.7623  & 27.72  & 0.7659  & \textcolor{red}{\textbf{27.83}}  & \textcolor{red}{\textbf{0.7692}}  \\
    D-Residential & 23.45  & 0.6301  & 24.46  & 0.6922  & 24.60  & 0.7004  & 24.64  & 0.7026  & \textcolor{blue}{\textbf{24.69}}  & \textcolor{blue}{\textbf{0.7051}}  & 24.56  & 0.6985  & 24.60  & 0.7029  & \textcolor{red}{\textbf{24.72}}  & \textcolor{red}{\textbf{0.7067}}  \\
    Desert & 37.14  & 0.8869  & 37.54  & 0.8933  & 37.58  & 0.8944  & \textcolor{blue}{\textbf{37.63}}  & \textcolor{blue}{\textbf{0.8951}}  & 37.62  & 0.8949  & 37.58  & 0.8941  & 37.60  & 0.8946  & \textcolor{red}{\textbf{37.65}}  & \textcolor{red}{\textbf{0.8954}}  \\
    Farmland & 32.32  & 0.7934  & 33.18  & 0.8166  & 33.33  & 0.8212  & 33.35  & 0.8216  & \textcolor{blue}{\textbf{33.42}}  & \textcolor{blue}{\textbf{0.8236}}  & 33.30  & 0.8202  & 33.39  & 0.8235  & \textcolor{red}{\textbf{33.45}}  & \textcolor{red}{\textbf{0.8245}}  \\
    Forest & 27.67  & 0.6411  & 27.84  & 0.6514  & 27.90  & 0.6558  & \textcolor{blue}{\textbf{27.95}}  & \textcolor{blue}{\textbf{0.6605}}  & 27.95  & 0.6604  & 27.88  & 0.6552  & 27.92  & 0.6598  & \textcolor{red}{\textbf{27.99}}  & \textcolor{red}{\textbf{0.6617}}  \\
    Industrial & 26.29  & 0.7063  & 27.43  & 0.7566  & 27.62  & 0.7647  & 27.67  & 0.7669  & \textcolor{blue}{\textbf{27.71}}  & \textcolor{blue}{\textbf{0.7689}}  & 27.56  & 0.7621  & 27.62  & 0.7665  & \textcolor{red}{\textbf{27.74}}  & \textcolor{red}{\textbf{0.7701}}  \\
    Meadow & 32.07  & 0.7100  & 32.33  & 0.7136  & 32.38  & 0.7157  & 32.40  & 0.7167  & \textcolor{blue}{\textbf{32.41}}  & 0.7172  & 32.37  & 0.7154  & 32.39  & \textcolor{blue}{\textbf{0.7175}}  & \textcolor{red}{\textbf{32.43}}  & \textcolor{red}{\textbf{0.7176}}  \\
    M-Residential & 26.32  & 0.6578  & 27.40  & 0.7055  & 27.55  & 0.7124  & 27.58  & 0.7136  & \textcolor{blue}{\textbf{27.63}}  & \textcolor{blue}{\textbf{0.7159}}  & 27.52  & 0.7113  & 27.55  & 0.7140  & \textcolor{red}{\textbf{27.66}}  & \textcolor{red}{\textbf{0.7170}}  \\
    Mountain & 28.05  & 0.6978  & 28.41  & 0.7127  & 28.48  & 0.7163  & \textcolor{blue}{\textbf{28.50}}  & \textcolor{blue}{\textbf{0.7184}}  & 28.50  & 0.7183  & 28.46  & 0.7152  & 28.45  & 0.7171  & \textcolor{red}{\textbf{28.52}}  & \textcolor{red}{\textbf{0.7192}}  \\
    Park  & 27.14  & 0.7060  & 27.77  & 0.7332  & 27.87  & 0.7383  & 27.91  & 0.7406  & \textcolor{blue}{\textbf{27.93}}  & \textcolor{blue}{\textbf{0.7415}}  & 27.83  & 0.7365  & 27.86  & 0.7395  & \textcolor{red}{\textbf{27.96}}  & \textcolor{red}{\textbf{0.7431}}  \\
    Parking & 24.11  & 0.7310  & 26.18  & 0.8059  & 26.51  & 0.8159  & 26.68  & 0.8205  & \textcolor{blue}{\textbf{26.76}}  & 0.8225  & 26.46  & 0.8145  & 26.70  & \textcolor{blue}{\textbf{0.8226}}  & \textcolor{red}{\textbf{26.81}}  & \textcolor{red}{\textbf{0.8238}}  \\
    Playground & 29.33  & 0.7723  & 30.70  & 0.8132  & 30.94  & 0.8200  & 31.03  & 0.8216  & \textcolor{blue}{\textbf{31.10}}  & \textcolor{blue}{\textbf{0.8235}}  & 30.88  & 0.8182  & 31.03  & 0.8230  & \textcolor{red}{\textbf{31.13}}  & \textcolor{red}{\textbf{0.8244}}  \\
    Pond  & 28.76  & 0.7559  & 29.38  & 0.7749  & 29.46  & 0.7781  & 29.49  & 0.7790  & \textcolor{blue}{\textbf{29.51}}  & \textcolor{blue}{\textbf{0.7796}}  & 29.44  & 0.7769  & 29.46  & 0.7788  & \textcolor{red}{\textbf{29.53}}  & \textcolor{red}{\textbf{0.7802}}  \\
    Port  & 25.83  & 0.7727  & 26.94  & 0.8207  & 27.11  & 0.8255  & 27.18  & 0.8273  & \textcolor{blue}{\textbf{27.21}}  & \textcolor{red}{\textbf{0.8283}}  & 27.05  & 0.8237  & 27.13  & 0.8267  & \textcolor{red}{\textbf{27.22}}  & \textcolor{blue}{\textbf{0.8280}}  \\
    Railway Station & 27.00  & 0.6965  & 27.87  & 0.7347  & 28.04  & 0.7434  & 28.11  & 0.7456  & \textcolor{blue}{\textbf{28.13}}  & \textcolor{blue}{\textbf{0.7474}}  & 27.99  & 0.7404  & 28.07  & 0.7461  & \textcolor{red}{\textbf{28.16}}  & \textcolor{red}{\textbf{0.7489}}  \\
    Resort & 26.78  & 0.7160  & 27.67  & 0.7547  & 27.81  & 0.7601  & 27.86  & 0.7622  & \textcolor{blue}{\textbf{27.89}}  & \textcolor{blue}{\textbf{0.7634}}  & 27.78  & 0.7588  & 27.82  & 0.7619  & \textcolor{red}{\textbf{27.92}}  & \textcolor{red}{\textbf{0.7644}}  \\
    River & 29.29  & 0.7201  & 29.74  & 0.7378  & 29.81  & 0.7412  & 29.84  & 0.7426  & \textcolor{blue}{\textbf{29.85}}  & \textcolor{blue}{\textbf{0.7430}}  & 29.79  & 0.7401  & 29.80  & 0.7422  & \textcolor{red}{\textbf{29.88}}  & \textcolor{red}{\textbf{0.7439}}  \\
    School & 25.81  & 0.7046  & 26.81  & 0.7503  & 26.96  & 0.7568  & 27.00  & 0.7587  & \textcolor{blue}{\textbf{27.04}}  & \textcolor{blue}{\textbf{0.7604}}  & 26.91  & 0.7546  & 26.94  & 0.7577  & \textcolor{red}{\textbf{27.07}}  & \textcolor{red}{\textbf{0.7619}}  \\
    S-Residential & 25.71  & 0.5790  & 26.15  & 0.6001  & 26.22  & 0.6055  & 26.25  & 0.6070  & \textcolor{blue}{\textbf{26.27}}  & \textcolor{blue}{\textbf{0.6079}}  & 26.21  & 0.6046  & 26.23  & 0.6075  & \textcolor{red}{\textbf{26.29}}  & \textcolor{red}{\textbf{0.6080}}  \\
    Square & 27.51  & 0.7286  & 28.71  & 0.7782  & 28.90  & 0.7847  & 28.96  & 0.7870  & \textcolor{blue}{\textbf{28.99}}  & \textcolor{blue}{\textbf{0.7881}}  & 28.85  & 0.7831  & 28.92  & 0.7868  & \textcolor{red}{\textbf{29.03}}  & \textcolor{red}{\textbf{0.7891}}  \\
    Stadium & 26.23  & 0.7241  & 27.55  & 0.7771  & 27.75  & 0.7845  & 27.82  & 0.7873  & \textcolor{blue}{\textbf{27.85}}  & \textcolor{blue}{\textbf{0.7882}}  & 27.68  & 0.7823  & 27.78  & 0.7867  & \textcolor{red}{\textbf{27.87}}  & \textcolor{red}{\textbf{0.7889}}  \\
    Storage Tanks & 25.15  & 0.6667  & 26.25  & 0.7190  & 26.39  & 0.7255  & 26.41  & 0.7264  & \textcolor{blue}{\textbf{26.45}}  & \textcolor{blue}{\textbf{0.7287}}  & 26.33  & 0.7233  & 26.39  & 0.7274  & \textcolor{red}{\textbf{26.47}}  & \textcolor{red}{\textbf{0.7298}}  \\
    Viaduct & 26.77  & 0.6817  & 27.70  & 0.7225  & 27.88  & 0.7310  & 27.91  & 0.7325  & \textcolor{blue}{\textbf{27.98}}  & \textcolor{blue}{\textbf{0.7361}}  & 27.81  & 0.7277  & 27.87  & 0.7326  & \textcolor{red}{\textbf{28.00}}  & \textcolor{red}{\textbf{0.7378}}  \\
\midrule
\midrule
    Average & 27.89  & 0.7238  & 28.80  & 0.7597  & 28.94  & 0.7653  & 28.99  & 0.7671  & \textcolor{blue}{\textbf{29.02}}  & \textcolor{blue}{\textbf{0.7684}}  & 28.90  & 0.7637  & 28.96  & 0.7672  & \textcolor{red}{\textbf{29.05}}  & \textcolor{red}{\textbf{0.7694}}  \\
    \bottomrule[1.5pt]
    \end{tabular}%
    }}
  \label{aid-res}%
\end{table*}%

\subsection{Comparisons With State-of-the-Art}

\subsubsection{Comparative Methods}

CNN-based models, including SRCNN~\cite{srcnn}, EDSR~\cite{edsr}, and HAUNet~\cite{HAUNet}; GAN-based models, including SRGAN~\cite{srgan}; Transformer-based models, including SwinIR~\cite{swinir}, HAT~\cite{hat}, and SPT~\cite{SPT}; CNN-Transformer hybrid models, including ConvFormerSR~\cite{ConvFormerSR}; and Mamba-based models, including MambaIR~\cite{mambair} and FreMamba~\cite{FreMamba}, were involved in evaluating the SR performance of our GDSR against SOTA methods on RSI-SR.

\begin{figure*}[!t]
\centering
\includegraphics[width=7in]{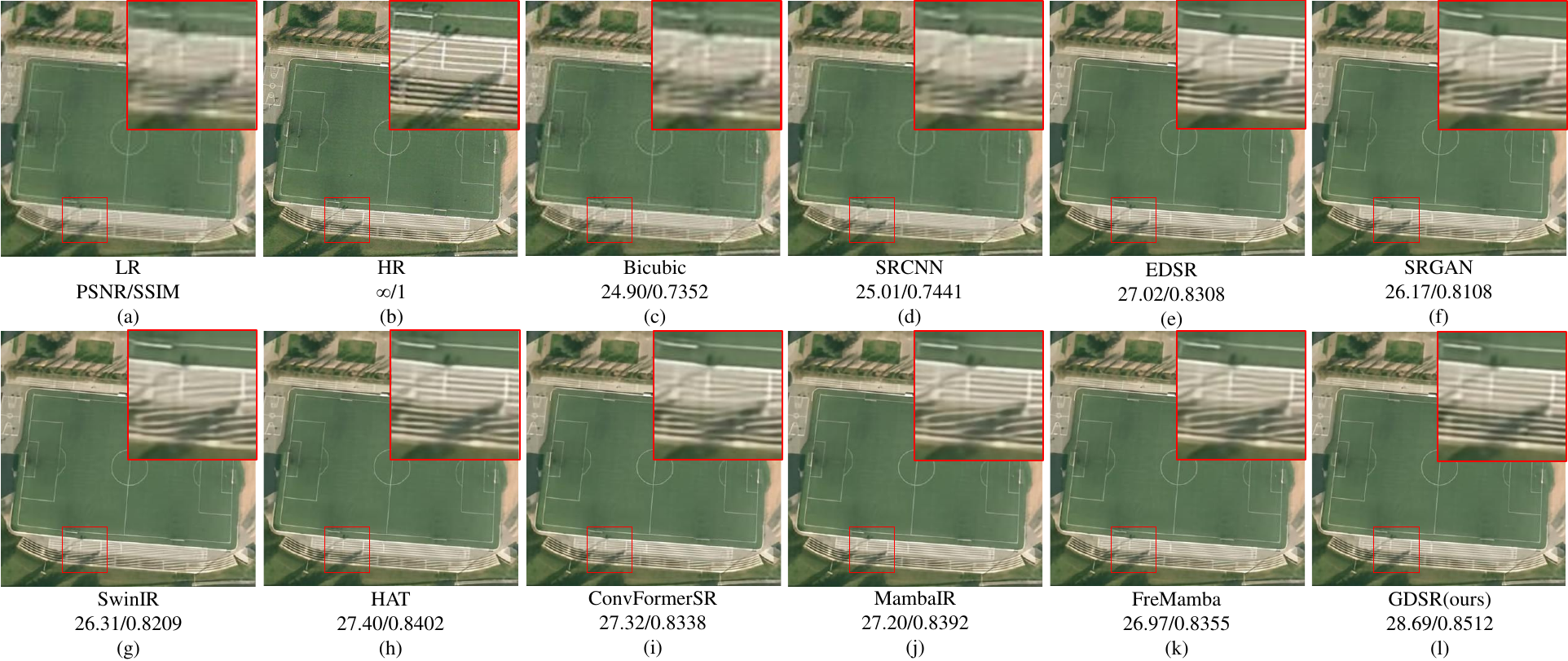}%
\captionsetup{font={scriptsize}}
\caption{Visual comparisons of our GDSR with CNN-, GAN-, Transformer-, and Mamba-based methods on the AID dataset's "playground\_161" image with scale ×3. Zoom in for better observation.}
\label{aid-vis}
\end{figure*}

\begin{figure*}[!t]
\centering
\includegraphics[width=7in]{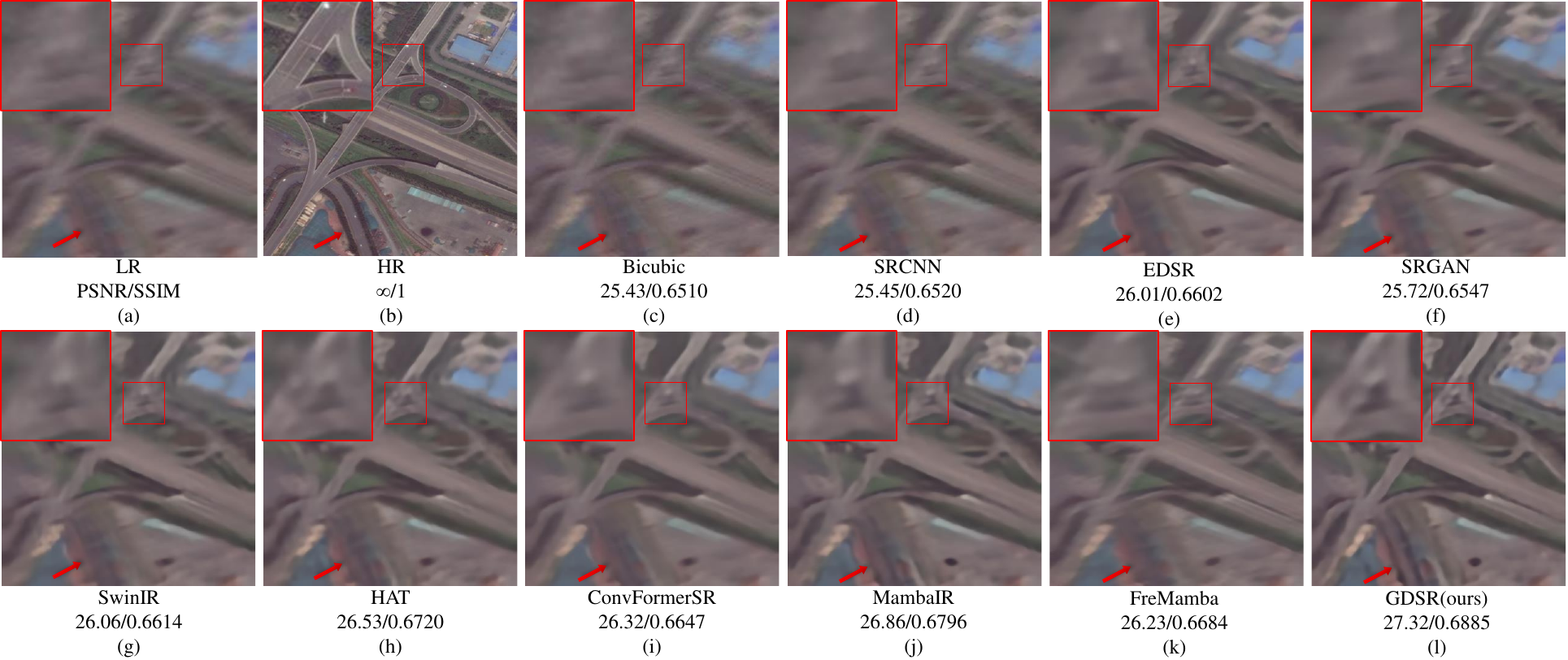}%
\captionsetup{font={scriptsize}}
\caption{Visual comparisons of our GDSR with CNN-, GAN-, Transformer-, and Mamba-based methods on the AID\_CDM dataset's "viaduct\_398" image with scale ×3. Zoom in for better observation.}
\label{aid_CDM_vis}
\end{figure*}

\subsubsection{Quantitative Evaluations}

The quantitative evaluation results across different datasets and scaling factors were comprehensively analyzed through Tables~\ref{all-res} and~\ref{qh-res} to assess reconstruction accuracy and perceptual quality. On the AID dataset under bicubic degradation, the proposed GDSR demonstrated state-of-the-art performance with PSNR values of 31.81 dB (×2), 29.05 dB (×3), and 27.39 dB (×4), showing improvements of 0.04-0.11 dB over transformer-based HAT and CNN-based EDSR. The structural preservation capability was further verified by SSIM metrics, where GDSR achieved 0.8592 (×2), 0.7694 (×3), and 0.7022 (×4). Although FreMamba exhibited superior LPIPS scores across ×2-4 scales, GDSR maintained competitive perceptual quality while balancing pixel-wise accuracy and visual realism. For more challenging CDM-synthesized data, GDSR showed remarkable robustness with consistent performance gains, surpassing MambaIR by 0.11 dB PSNR and 0.26\% SSIM in ×2 super-resolution. The superiority became more evident on the UCMerced dataset, where GDSR established new benchmarks in 7 out of 12 metrics. Notably, it achieved 0.17 dB PSNR improvement over MambaIR and 0.11 dB gain against HAT in ×3 super-resolution under CDM degradation. Evaluation on the real-world RSSRD-QH dataset further validated the practical effectiveness of GDSR. Under bicubic degradation, the method attained peak PSNR values of 29.87 dB (×2), 26.73 dB (×3), and 25.12 dB (×4). In CDM degradation scenarios, GDSR exhibited particularly strong performance at high super-resolution factors, reaching 22.87 dB PSNR (×3) and 22.27 dB PSNR (×4) with corresponding SSIM values of 0.3864 and 0.3488.

To further evaluate the generalization capability of the super-resolution models across diverse remote sensing scenarios, we report the PSNR and SSIM results across 30 scene categories in the AID dataset under bicubic degradation, as shown in Table~\ref{aid-res}. As evidenced, GDSR exhibited stronger generalization capability compared to state-of-the-art methods, achieving the best performance in nearly all remote sensing scenarios. Specifically, for the "Parking" and "Playground" scenes, GDSR outperformed HAT by 0.13 dB and 0.1 dB in PSNR and by 0.33\% and 0.28\% in SSIM, respectively. Additionally, in the most complex categories, such as "Church" and "DenseResidential," GDSR also achieved the best performance.

\begin{figure*}[!t]
\centering
\includegraphics[width=7in]{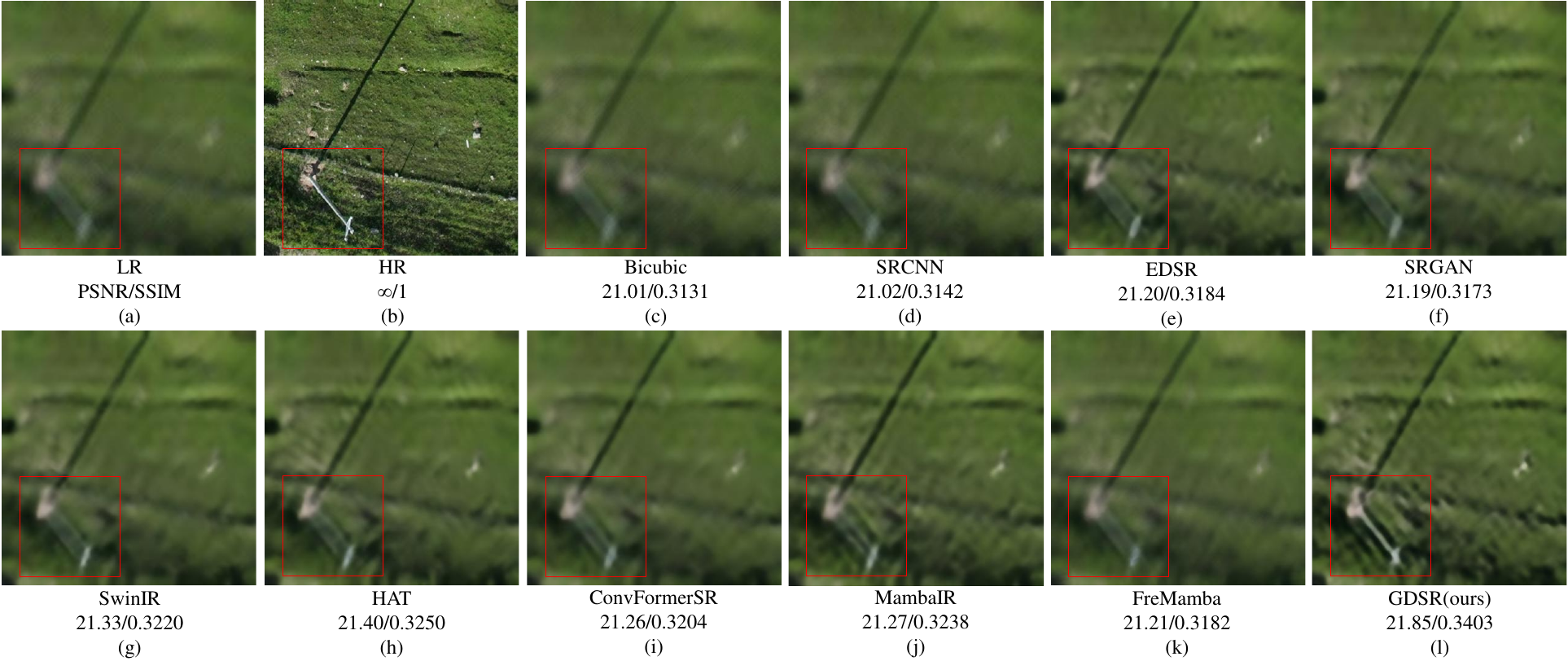}%
\captionsetup{font={scriptsize}}
\caption{Visual comparisons of our GDSR with CNN-, GAN-, Transformer-, and Mamba-based methods on the RSSRD-QH\_CDM dataset with scale ×3.}
\label{QH_CDM_vis}
\end{figure*}

Moreover, as the receptive field of the models increased, performance trended upward for CNN-based EDSR, Transformer-based SwinIR, and Mamba-based MambaIR. This aligned with our experimental results on the RGEG Branch and RDEG Branch, where larger receptive fields enabled the activation of more pixels for image reconstruction, enhancing performance. However, the results from MambaIR and FreMamba, along with the experiments on F2B and F2C, indicated that spatial global modeling in complex RSIs had reached a bottleneck. Simply increasing the receptive field no longer yielded significant performance gains. A more practical solution was to effectively utilize large receptive fields to activate the appropriate regions and pixels for accurate image reconstruction. The outstanding generalization capability and superior performance of GDSR align with our motivation to address the challenges of large-scale RSIs efficiently through the introduction of a global-detail dual-branch structure.

\subsubsection{Complexity and Efficiency Evaluation}

\begin{figure}[!t]
\centering
\includegraphics[width=3.4in]{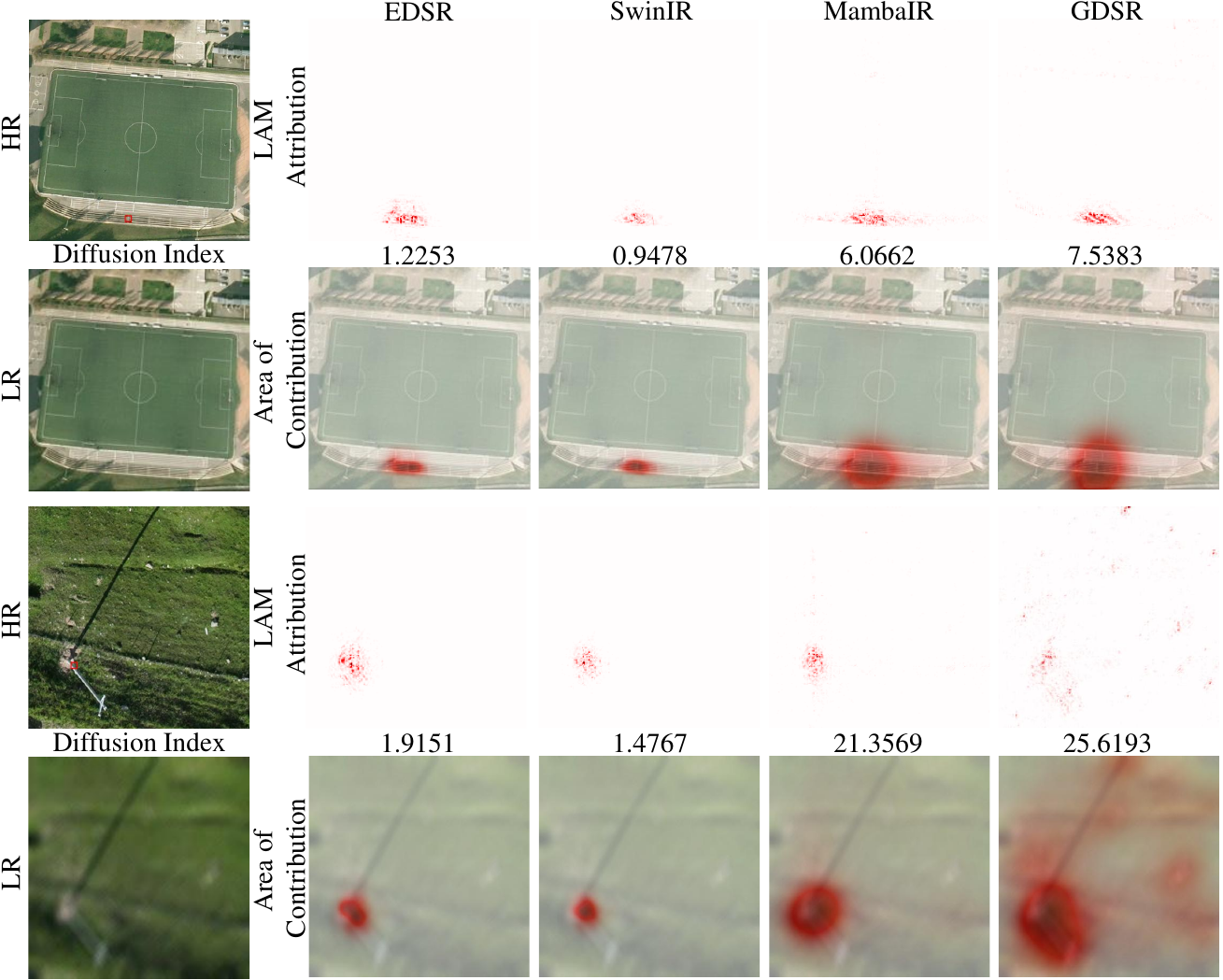}%
\captionsetup{font={scriptsize}}
\caption{The visualization of LAM on datasets AID and RSSRD-QH\_CDM.}
\label{LAM_vis}
\end{figure}

To evaluate the complexity and computational efficiency of our proposed models, we conducted quantitative assessments across various metrics. As shown in Table~\ref{complexity}, the bicubic interpolation technique inherently avoids the need for parameter optimization, resulting in the fastest inference speed within our evaluation scope. The CNN-based model achieved the highest inference speed among all tested deep learning models, primarily due to the inherent efficiency of convolutional parallel operations. The EDSR model, with its deep residual structure, had the largest number of parameters and FLOPs, yet achieved inference speeds several times faster than other architectures.

\begin{table}[!t]
  \centering
  \captionsetup{font={scriptsize}, labelsep = newline, justification=centering}
  \caption{Comparisons of model complexity and efficiency.}
    \begin{tabular}{ccccc}
    \toprule[1.5pt]
    Model & \#Param. & FLOPs & Memory & FPS\\
    \midrule    
    \midrule
    Bicubic     &  \emph{-}     & \emph{-} & 22MB & 27498.0 \\
    SRCNN~\cite{srcnn}     &  0.02M     & 4.65G & 140MB & 1109.0 \\
    SRGAN~\cite{srgan}     &  1.55M     & 47.86G & 300MB & 323.3 \\
    EDSR~\cite{edsr}     &  43.68M     & 1119.98G & 1068MB & 35.1 \\
    SwinIR~\cite{swinir}     &  16.62M     & 446.67G & 516MB & 5.5 \\
    HAT~\cite{hat}     &  20.81M     & 652.17G & 1446MB & 4.6 \\
    HAUNet~\cite{HAUNet}     &  2.64M     & 1353.07G & 7658MB & 3.1 \\
    SPT~\cite{SPT}     &  3.25M     & 668.46G & 9974MB & 2.2 \\
    ConvFormerSR~\cite{ConvFormerSR}     &  16.49M     & 442.18G & 500MB & 7.0 \\
    MambaIR~\cite{mambair}     &  15.11M     & 372.85G & 556MB & 6.4 \\
    FreMamba~\cite{FreMamba}     &  11.46M     & 229.50G & 546MB & 5.5 \\
    \hline
    \hline
    \rowcolor{mygray}\textbf{GDSR (Ours)} &  13.17M     & 338.73G & 518MB & 14.8 \\
    \bottomrule[1.5pt]
    \end{tabular}%
  \label{complexity}%
\end{table}%

\begin{figure*}[!t]
\centering
\includegraphics[width=7in]{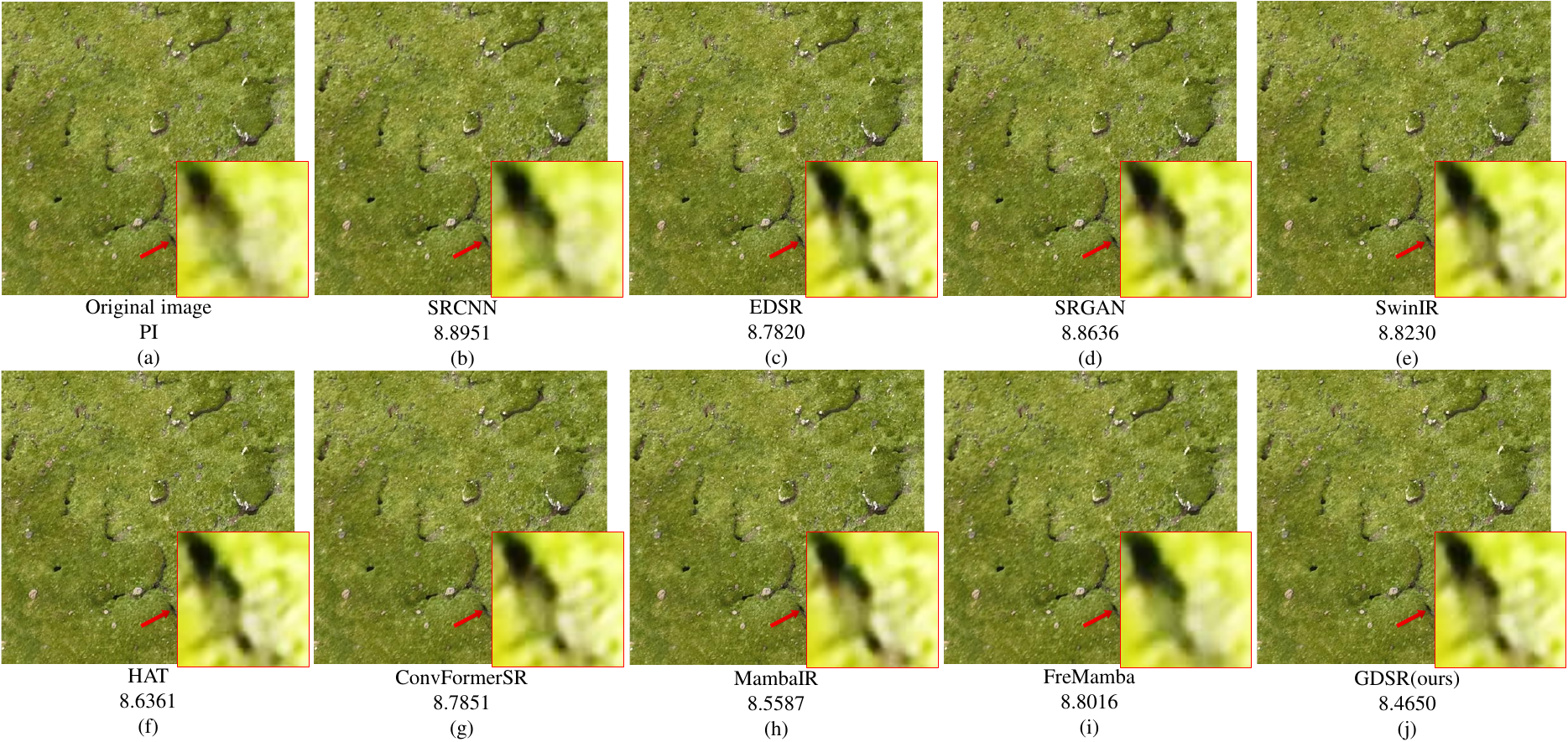}%
\captionsetup{font={scriptsize}}
\caption{Visualizations of Real-World Images for better observation of mouse holes. Zoom in for better observation. We tested the model trained with RSSRD-QH\_CDM on real drone images. A lower value indicates better reconstruction quality when using the Perception Index (PI) metric~\cite{blau20182018}. Enhanced brightness and contrast for better observation.}
\label{real_vis}
\end{figure*}

Transformer-based models, such as SwinIR and HAT, exhibited slower inference speeds compared to CNN-based models but achieved significant performance gains. Models based on Mamba, including MambaIR and FreMamba, leveraged their linear complexity and optimized inference processes, delivering superior performance along with relatively faster inference speeds. ConvFormerSR, which integrated Swin Transformer with CNN architectures, achieved faster inference speeds while maintaining competitive performance. GDSR, by incorporating RWKV with CNN, not only reduced the parameter count but also achieved inference speeds several times faster than Transformer- and Mamba-based models, while delivering enhanced performance. As demonstrated in Tables~\ref{num-RCB-R-RWKVB},~\ref{GDRM}, and~\ref{depth}, these results underscore the superiority of the Global-Detail dual-branch architecture design.

Comparing Tables~\ref{complexity} and~\ref{all-res} reveals that the complexity of our proposed model is comparable to most state-of-the-art models, while its efficiency is superior, and its performance is significantly better. These results highlight the practical potential of our model, demonstrating an effective balance between computational efficiency and performance, making it well-suited for real-world applications.

\subsubsection{Qualitative Results}
Visual comparisons on AID, AID\_CDM, and RSSRD-QH\_CDM with a scale factor of ×3 are presented in Fig.~\ref{aid-vis}, Fig.~\ref{aid_CDM_vis}, and Fig.~\ref{QH_CDM_vis}. From these visual results, it was evident that our proposed GDSR was capable of restoring sharp edges and showcasing richer textures, particularly in capturing critical high-frequency details in remote sensing images. For instance, when comparing the reconstructed "playground\_161" in Fig.~\ref{aid-vis}, the closest competing Transformer-based method HAT, the hybrid CNN-Transformer method ConvFormerSR, and the Mamba-based model MambaIR all failed to reconstruct the high-resolution lines on the runway. Only the proposed GDSR accurately restored these details. Furthermore, as illustrated in Fig.~\ref{aid_CDM_vis} for "viaduct\_398," under more complex degradation scenarios, GDSR effectively reconstructed the contours, whereas other SR models failed to accurately represent the detailed structure. In Fig.~\ref{QH_CDM_vis}, a comparison of the visual results for the RSSRD-QH\_CDM dataset showed that the original image, after CDM degradation, appeared heavily blurred and lost its distinctive columnar features. Only the proposed GDSR was able to accurately recover the outline of the original image.

The LAM comparisons in Fig.~\ref{LAM_vis}, based on Figs.~\ref{aid-vis} and~\ref{QH_CDM_vis}, demonstrated that GDSR benefited from the superior long-range modeling capability of RGEG, enabling it to leverage more pixels during the SR process. Additionally, the strong detail extraction capability of RDEG allowed GDSR to utilize pixels correctly for reconstruction. Specifically, for the RSSRD-QH\_CDM dataset, when large-scale and global information was present, CNN- and Transformer-based SR networks struggled to extract sufficient information, and while the Mamba-based SR network activated a large number of pixels, it failed to use them correctly for reconstruction. In contrast, thanks to the effective integration of RGEG and RDEG, the GDSR-generated results significantly outperformed other SR models. Moreover, GDSR surpassed MambaIR in terms of a diffusion index by 25.6193, indicating that by incorporating an RWKV with a larger receptive field into RGEG, GDSR could better capture global information in images with more such features, leading to a higher diffusion index. These visual comparisons further validated the effectiveness of GDSR in capturing and reconstructing details in remote sensing images, highlighting the powerful pixel utilization capability brought by our global-detail dual-branch structure.

\subsubsection{Real-World Image Testing on the Sanjiangyuan Region}

To further evaluate the model's performance in real-world scenarios, we tested it on RSIs from real-world environments. As shown in Fig.~\ref{real_vis}, the image depicts an alpine meadow grassland type, with surface features such as cracks, scattered rocks, and numerous small "mouse holes." These mouse holes are depressions or openings on the ground, and the distribution of rodents is significant for studying the primary grassland type, the alpine meadow, in Qinghai Province. Their presence can impact the biodiversity, productivity, and ecological function of the alpine meadows. However, due to the limitations in drone flight conditions and resolution, the raw images suffer from motion blur and loss of details. These issues hinder accurate observation of key features such as surface textures and the depth of the mouse holes. By applying SR techniques, our GDSR effectively enhanced the clarity of these fine details, sharpened the edges around the mouse holes, and restored the surface texture. This improvement is crucial for accurate environmental analysis, as the restored landscape provides more precise data for studying biodiversity conservation and the enhancement of ecological functions in the alpine meadow ecosystem of Qinghai Province.

\section{Discussion}

The proposed method demonstrated consistent SR performance across diverse remote sensing scenarios, validated on the AID (30 land-use categories)~\cite{aid}, UCMerced (21-class geospatial scenes)~\cite{UCMerced}, and custom RSSRD-QH (Qinghai Plateau ecology) datasets. By enhancing texture fidelity in complex environments such as fragmented agricultural landscapes, high-density urban settlements, and ecologically sensitive alpine regions, this approach supports critical applications requiring precise feature extraction. Improved road network reconstruction in severely degraded AID samples aids emergency routing during disasters, while sharpened landscape details in RSSRD-QH enable accurate environmental monitoring. The method’s adaptability to both global benchmarks and region-specific challenges positions it as a practical tool for revitalizing historical mid-resolution archives into high-precision analysis-ready data.

\begin{figure}[!t]
\centering
\includegraphics[width=3.4in]{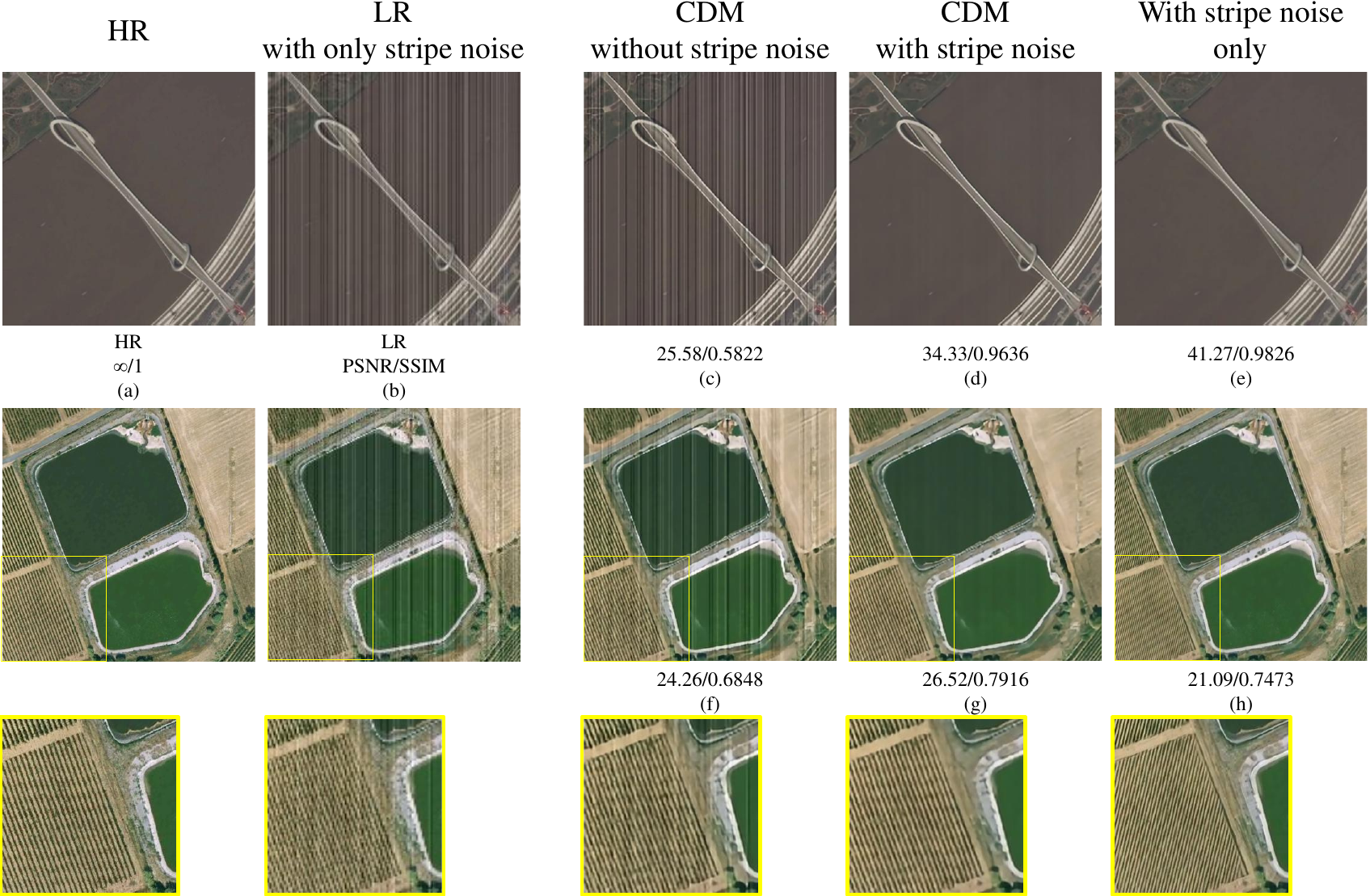}%
\captionsetup{font={scriptsize}}
\caption{Visual comparison of GDSR's adaptability to severe remote sensing-specific stripe noise on two samples from the AID dataset. The top and bottom rows show results for "bridge\_43" and "pond\_410", respectively. (c, f) Results from the original GDSR model (trained on CDM without stripe noise) failed to remove the stripes. (d, g) Results from the retrained GDSR model (trained on CDM with stripe noise) successfully removed the stripes and reconstructed the scene. (e, h) Results from the retrained model on data with only stripe noise, showing its targeted destriping capability.}
\label{rsinoise}
\end{figure}

Additionally, we tested GDSR's robustness against challenging, remote sensing-specific artifacts. While our initial CDM followed established blind SR methods~\cite{blind1,blind2,blind3,refdiff}, real-world RSIs often contain severe sensor-induced stripe noise. To validate our method’s adaptability, we augmented the CDM with this artifact and retrained the GDSR model. As demonstrated in Fig.~\ref{rsinoise}, the results confirmed GDSR's strong adaptive capabilities. The retrained model successfully identified and eliminated severe stripe noise in images, leading to quantitative and visual improvements. This positive result paves the way for future work to incorporate other challenging degradations, such as atmospheric scattering, further enhancing the model's practical utility.

\section{Conclusion} \label{section5}

In this study, we first introduce the RWKV for RSISR. GDSR effectively models the global dependencies of large-scale RSI while incorporating local details, all with linear complexity and significantly improved inference speed. Specifically, we design an efficient and effective Global-Detail dual-branch architecture that correctly activates more pixels spatially for SR. To combine global and local representations, we propose GDRM, which adaptively adjusts the features of the dual branches and fuses them. Additionally, we introduce Dual-Group Multi-Scale Wavelet Loss to help the model better reconstruct high-frequency details. Extensive quantitative and qualitative experiments on the AID, UCMerced, and RSSRD-QH benchmark datasets using two degradation methods demonstrate that our GDSR outperforms state-of-the-art CNN-, Transformer-, and Mamba-based SR models in RSI-SR tasks.

Nevertheless, our GDSR does exhibit some shortcomings. First, the use of simple stacked residual convolutions for detail extraction introduces a high parameter count, limiting model efficiency. Future work will explore more efficient architectures specifically tailored for RSIs. Second, the current validation is limited to RGB RSIs, lacking evaluation on hyperspectral SR scenarios where balancing spatial enhancement with spectral fidelity is paramount. Extending the method to hyperspectral data and rigorously assessing its spectral accuracy will be a key future direction. Moreover, we consider investigating the model's applicability to multi-temporal or cross-sensor RSIs, addressing challenges like temporal consistency and domain adaptation to further enhance its practical utility.

\ifCLASSOPTIONcaptionsoff
  \newpage
\fi

\bibliographystyle{IEEEtran}
\bibliography{reference}

\end{document}